\documentclass[preprint,showpacs,preprintnumbers,amsmath,amssymb,prb]{revtex4-1}

\usepackage[dvipdfmx]{graphicx}
\usepackage{dcolumn}
\usepackage{bm}
\usepackage{color} 
\usepackage{ulem} %
\begin{document}
%
\title{Transient photocurrent and optical absorption of disordered thin-film semiconductors: in-depth injection and nonlinear response}
\author{
Kazuhiko Seki
}
\email{k-seki@aist.go.jp}
\affiliation{National Institute of Advanced Industrial Science and Technology (AIST), Onogawa 16-1 AIST West, Ibaraki, 305-8569, Japan
}
\author{Naoya Muramatsu}
\affiliation{
Graduate School of Science and Technology, Niigata University, 2-8050, Ikarashi, Nishi-Ku, Niigata 950-2181, Japan }
\author{Tomoaki Miura}
\affiliation{
Graduate School of Science and Technology, Niigata University, 2-8050, Ikarashi, Nishi-Ku, Niigata 950-2181, Japan }
\author{Tadaaki Ikoma}
\affiliation{
Graduate School of Science and Technology, Niigata University, 2-8050, Ikarashi, Nishi-Ku, Niigata 950-2181, Japan }

\date{\today}
\begin{abstract}
The time-of-flight method is a fundamental approach for characterizing the transport properties of semiconductors. 
Recently, the transient photocurrent and optical absorption kinetics have been simultaneously measured for thin films; pulsed-light excitation of thin films should give rise to non-negligible in-depth carrier injection. 
Yet, the effects of in-depth carrier injection on the transient currents and optical absorption have not yet been elucidated theoretically. 
Here, by considering the in-depth carrier injection in simulations, we found a $1/t^{1-\alpha/2}$ initial time ($t$) dependence rather than the conventional $1/t^{1-\alpha}$ dependence under a weak external electric field, where $\alpha<1$ is the index of dispersive diffusion.
The asymptotic transient currents are not influenced by the initial in-depth carrier injection and follow the conventional $1/t^{1+\alpha}$ time dependence. 
We also present the relation between the field-dependent mobility coefficient and the diffusion coefficient when the transport is dispersive. 
The field dependence of the transport coefficients influences the transit time in the photocurrent kinetics dividing two power-law decay regimes. 
The classical Scher--Montroll theory predicts $a_1+a_2=2$ when the initial photocurrent decay is given by $1/t^{a_1}$ and the asymptotic photocurrent decay is given by $1/t^{a_2}$. 
The results shed light on the interpretation of the power-law exponent of $1/t^{a_1}$ when $a_1+a_2\neq 2$. 
\end{abstract}

\maketitle
\section{Introduction}
The time-of-flight (TOF) method is a fundamental means of characterizing the transport properties of semiconductors. 
In the TOF method, charge carriers in the vicinity of the transparent back electrode are excited by pulsed light and the transient photocurrent reflecting the transport of photo-generated charge carriers to the front electrode is measured; the transit time of charge carriers to the front electrode can then be determined. \cite{Scher_75,Pfister_78,Tiedje_81,Nebel_89} 
In general, pulsed-light excitation of thin films requires consideration of the depth of carrier injection. 
Conventional TOF measurements have been carried out for devices with a thick (on the order of $\sim1 \mu$m) semiconductor layer sandwiched between coplanar electrodes so that the photocarriers are generated only in the vicinity of the transparent electrode and not throughout the entire layer.  \cite{YOSHIKAWA_08,Kougo_16,Bratina_19} 
However, there is a need for TOF-type analysis methods for transient photocurrents of thin ($<100$ nm) film devices such as organic solar cells, where in-depth carrier injection should be considered. \cite{Kudo_18}
Recently, the transient photocurrent and optical absorption kinetics have been simultaneously measured for organic thin films to investigate the relation between the time change of the carrier number density and the photocurrent; this method is known as simultaneous optical and electrical detection (SOED). \cite{Miura_21} 
The SOED technique has been applied to organic solar cells, and preliminary results have been obtained. 
However, the time change of the carrier density due to drift/diffusion transport has not yet been elucidated theoretically using the same settings as the transient photocurrent. 
Therefore, it is desirable to obtain analytical results for the kinetics of photocarrier density and photocurrent by solving the same transport equation assuming the same boundary and initial conditions.
In-depth carrier injection can be also induced in TOF photocurrent measurements using coplanar electrodes and a thin layer of molecular solids on a substrate. \cite{YOSHIKAWA_08,Kougo_16,Bratina_19,Tyutnev_14} 
Therefore, the in-depth carrier injection should be considered when simulating TOF and optical absorption signals.

Researchers have developed theories on the transient photocurrent using a continuous time random walk (CTRW) model or a drift-diffusion equation approach (fractional Fokker--Planck equation for dispersive diffusion) by assuming a linear response. \cite{Scher_75,Hirao_95,Nishizawa_06,Barkai_01}
The assumption of a linear response limits the photocurrent under a weak electric field. 
Although the assumption of a linear response might be satisfactory for thick layers, the field strength can exceed the limit of the applicability of the linear response when the film is thin under the same voltage difference.
We generalize the CTRW model applicable to a nonlinear response. 

For normal diffusion, where the position dispersion (variance) increases linearly with time, the transient photocurrent has been studied using the drift-diffusion equation under the condition of a constant applied voltage. \cite{Hirao_95,Nishizawa_06}
An analytical expression has been derived by assuming a constant electrostatic potential difference using the solution of the drift-diffusion equation in infinite space without imposing a boundary condition to represent charge extraction from the sample to the front electrode. 
Recently, the effect of the boundary conditions at both the front and back contacts to the respective electrodes has been rigorously considered. \cite{TYUTNEV_15}

For dispersive diffusion, where the position dispersion (variance) increases sub-linearly with time, the CTRW model has been used. 
When studying the transient photocurrent using the CTRW model, Scher and Montroll considered an absorbing boundary condition at the front contact with the electrode in their original analysis. \cite{Scher_75,Pfister_78} In their approach, however, the potential difference between the electrodes is not assumed to be a constant; a constant applied voltage has been rigorously considered only recently. \cite{Philippa_11,Sagues_17} 
Here, we study the transient photocurrent and optical absorption under a nonlinear response using the CTRW model by imposing rigorously the condition of a constant applied voltage. 

A method to analyze the transient photocurrents measured by the TOF method has been established using the CTRW approach for thick amorphous semiconductor layers. \cite{Scher_75,Pfister_78}
The TOF method has been used to extract the characteristic energy of disorder in the amorphous semiconductors through analysis of the log--log plot of the current against time after pulsed excitation of charge carriers in the vicinity of a transparent electrode. \cite{Scher_75,Pfister_78,Tiedje_81}
The characteristic energy of the exponential density of states can be obtained via the TOF method when a distinct power-law decay is observed before and after the transit time. 
In the presence of an exponential density of states, the charge-carrier transport becomes dispersive without a clear characteristic time scale for hopping transport; the photocurrent decays in the absence of a clear crossover unless the photocurrent is plotted against time using a log--log scale. \cite{Scher_75,Pfister_78} 

The TOF method has also been used to simultaneously obtain the diffusion constant and the mobility of thick molecularly doped polymers that exhibit normal diffusion rather than the dispersive diffusion. \cite{Hirao_95,Nishizawa_06}
Under a sufficiently strong electric field, the Einstein relation between the diffusion constant and the mobility is broken and simultaneous determination of these parameters is required under the constant potential difference between the electrodes. 
Recently, theoretical results for the TOF method have been advanced to satisfy the condition of a constant applied voltage even for dispersive diffusion as long as charge carriers are injected in the vicinity of the transparent electrode by pulsed light. \cite{Philippa_11,Sagues_17}
However, in the case where the thin-film semiconductors are amorphous and their energetic disorder is characterized by the exponential density of states, a theoretical interpretation has not yet been established for in-depth carrier injection.

In the present work, we use the CTRW model to develop an analytical theory for the transient photocurrent and optical absorption of amorphous thin-film semiconductors corresponding to in-depth carrier injection under various electric field strengths. 
The CTRW model enables us to study the electric field dependence of the dispersive currents and the dispersive diffusion. 
We show that the field dependence of the transport coefficients influences the transit time in the photocurrent kinetics dividing two power-law decay regimes.

\section{General formulation}

We first summarize the Scher--Montroll theory of the TOF method. \cite{Scher_75,Pfister_78}
In the Scher--Montroll theory, the transient current measured by the TOF method can be described by a power-law decay before the transit time and by another power law decay with a different exponent after the transit time.
We denote the magnitude of the exponent in the early time region by $a_1$ and that of the other exponent in the later time region by $a_2$. 
The power laws reflect the dispersive currents because of various time scales for detrapping of carriers from trap states; the absence of a specific time scale for detrapping originates from the exponential tail states. 
When the distribution of the trap states is expressed by $\exp \left(-E/E_0\right)/E_0$, where $E_0$ denotes the characteristic depth of the exponential trap energy distribution, $a_1=1-\alpha$ and $a_2=1+\alpha$ are derived from the CTRW model in the Scher--Montroll theory.
Here, $\alpha$ is given by $\alpha=k_{\rm B} T/E_0$, where $k_{\rm B}$ and $T$ are the Boltzmann constant and the temperature, respectively; they obey the relation $a_1+a_2=2$. 
 
We generalize the Scher--Montroll theory to include the initial charge-carrier distribution and to take into account the rigorous definition of the transient current when the static electrical potential is maintained as a constant. 
The CTRW is specified by the waiting-time distribution, $\psi (t)$, for detrapping from a trap state  
during $t$ and $t+dt$.
The power-law asymptotic tails in the waiting-time distribution result from an exponential density of states. 
The distribution function characterizing the density of states can be expressed as 
\begin{eqnarray}
g(E)= \exp \left( - E/E_0 \right) /E_0 .
\label{traped}
\end{eqnarray}
If the detrapping rate constant obeys the Arrhenius form,
\begin{eqnarray}
\gamma (E) = \gamma_{\rm r} \exp \left[ - E/(k_{\rm B} T) \right], 
\label{releaser}
\end{eqnarray}
then the waiting-time distribution function for detrapping is obtained as  \cite{Schnorer_88,Jakobs_93,Barzykin_02,Seki_03_1,Seki_03_2} 
\begin{align}
\psi (t) &= \int_0^{\infty} d\,E g(E) \gamma (E) \exp \left( - \gamma (E) t \right) 
\label{eq:psit}\\
&=\frac{\alpha \gamma \left( \alpha + 1, \gamma_{\rm r} t\right)}{\gamma_{\rm r}^\alpha t^{\alpha+1}} 
\sim \frac{\alpha \Gamma \left( \alpha + 1\right)}{\gamma_{\rm r}^\alpha t^{\alpha+1}} , 
\label{eq:psit_1}
\end{align}
where $\alpha \equiv k_{\rm B}T/E_0$, $\gamma (z, p) \equiv \int_0^p e^{-t} t^{z-1} d\,t \mbox{  for } (\mbox{Re} z > 0)$ is the incomplete Gamma function, $\Gamma (z)$ is the Gamma function, \cite{NIST} and $\gamma_{\rm r}$ is the elementary hopping rate constant; $\alpha <1$ is for dispersive transport. 
By introducing the Laplace transform
\begin{eqnarray}
\int_0^\infty dt\, \exp(-st) \psi (t) = \int_0^{\infty} d\,E g(E) \gamma (E) /\left[s+\gamma (E)\right] , 
\label{eq:Laplace}
\end{eqnarray}
we can express the waiting-time distribution function in the Laplace domain in the small $s$ limit as \cite{Barzykin_02,Seki_03_2}
\begin{align}
\hat{\psi }(s) 
=& 1 - \,_2F_1 \left[1, \alpha, \alpha+1, - \gamma_{\rm r} /s \right] 
\label{eq:psis}
\\
\sim& 1-\frac{\pi \alpha}{\sin \pi \alpha} \left( \frac{s}{\gamma_{\rm r}} \right)^\alpha ,
\label{eq:psis1}
\end{align}
where the integral representation of a hypergeometric function, $\,_2F_1 \left[1, \alpha, \alpha+1, - \gamma_{\rm r} /s \right] =\alpha \int_0^1 d\, y y^{\alpha-1}/(1+\gamma_{\rm r} y/s)$, is used. \cite{NIST} 
Here and below, we denote $\hat{f}$ as the Laplace transform of $f$.  
According to Tauberian theory, the Laplace transform of $t^{-\rho}$ is given by $\Gamma(1-\rho) s^{\rho-1}$; therefore, Eqs. (\ref{eq:psit}) and (\ref{eq:psis}) can be shown to be consistent with each other by noting that $\Gamma(1-\alpha)= -\alpha \Gamma(-\alpha)$ and $\Gamma(1-\alpha)\Gamma(1+\alpha)=\pi \alpha/\sin \pi \alpha$.

In the above discussion, we did not consider the effect of the external electric field on the hopping-rate constant. 
We refer to the largest hopping-rate constant [$E=0$ limit of Eq. (\ref{releaser})] as the elementary hopping-rate constant. 
Under the presence of an external electric field applied in the direction of one-dimensional hopping transitions, the elementary hopping-rate constant corresponding to the direction of increasing field strength (i.e., the forward direction) and that corresponding to the direction of decreasing field strength (i.e., the backward direction) are influenced by the external electric field strength. 
We consider the situation where an external electric field with strength $F$ is applied in the direction toward the extracting boundary located at $L$ along the $x$-axis. 
The location of the back-contact of the transparent electrode is denoted by $0$ along the $x$-axis.  
The elementary hopping-rate constant corresponding to the direction of the extracting boundary is denoted by $\gamma_{\rm rp}$, and the elementary hopping-rate constant  corresponding to the opposite direction is denoted by $\gamma_{\rm rm}$. 
Under the assumption that the Arrhenius law applies, these two elementary hopping-rate constants are given by 
\begin{align}
\gamma_{\rm rp}(F)&=(\gamma_{\rm r} /2)\exp\left[ q Fb/(2k_{\rm B} T)\right] 
\label{eq:gammap}
\\
\gamma_{\rm rm}(F)&=(\gamma_{\rm r}/2) \exp\left[ -q Fb/(2k_{\rm B} T)\right] ,
 \label{eq:gammam}  
 \end{align}
where $q$ is the elementary charge and $b$ is the hopping distance. 
The detailed balance condition can be confirmed, {\it i.e.}, $\gamma_{\rm rp}(F)/\gamma_{\rm rm}(F)=\exp\left[ q Fb/(k_{\rm B} T)\right] $. 
The factor two in $(\gamma_{\rm r}/2)$ is introduced because $\gamma_{\rm r}$ is the elementary hopping rate constant from a trap 
to both directions in one-dimensional transitions in the absence of an applied field. 
The detrapping rate given by Eq. (\ref{releaser}) becomes field dependent; the detrapping rate constant in the direction of the applied field is given by $\gamma_{\rm p} (E)= \gamma_{\rm rp} (F)\exp \left[ - E/(k_{\rm B} T) \right] $, and the detrapping rate constant in the direction opposite to the applied field is given by $\gamma_{\rm m} (E)= \gamma_{\rm rm} (F)\exp \left[ - E/(k_{\rm B} T) \right] $, where $E$ indicates the energy depth of the trap state in Eq. (\ref{releaser}).  
As a result, the total hopping frequency changes from $\gamma_{\rm r}$ to 
\begin{align}
\gamma_{\rm rt}(F)=\gamma_{\rm rp} (F)+\gamma_{\rm rm} (F)=\gamma_{\rm r}  \cosh\left[ q Fb/(2k_{\rm B} T)\right] 
\label{eq:gamma_rt},
\end{align}
which reduces to $\gamma_r$ in the limit of $F=0$.
By defining the total detrapping rate constant obeying an Arrhenius-type temperature dependence as
\begin{align}
\gamma_{\rm t} (E) =\gamma_{\rm rt}(F)\exp \left[ - E/(k_{\rm B} T) \right] ,
\label{releasert}
\end{align}
we can define the total waiting-time distribution under the bias by
\begin{align}
\psi_{\rm t} (t) = 
\int_0^{\infty} d\,E g(E) \gamma_{\rm t} (E)  \exp \left( - \gamma_{\rm t} (E)  t \right) .
\label{eq:psitt}
\end{align}
We ignore coupling between field strength and trap energy, which might influence the transient current at high electric field. 
\cite{Tachiya_10}
The waiting-time distribution along the direction of the extracting boundary and that along the opposite direction are given by  
$\psi_{\rm p} (t) =\Gamma_{\rm p} (F) \psi_{\rm t} (t) $  and $\psi_{\rm m} (t) =\Gamma_{\rm m} (F) \psi_{\rm t} (t) $, respectively.
We introduce the fraction of transitions from $x$ to $x+b$ among the sum of the transitions to $x+b$ and those to $x-b$ as
\begin{align}
\Gamma_{\rm p} (F) &= \frac{\gamma_{\rm rp}(F)}{\gamma_{\rm rt}(F)} =\frac{\exp\left[ q Fb/(2k_{\rm B} T)\right]}{2\cosh\left[ q Fb/(2k_{\rm B} T)\right]}
\label{eq:psip}
\end{align}
and the similarly defined fraction in the opposite direction as
\begin{align}
\Gamma_{\rm m} (F)&=  \frac{\gamma_{\rm rm}(F)}{\gamma_{\rm rt}(F)} =\frac{\exp\left[ -q Fb/(2k_{\rm B} T)\right]}{2\cosh\left[ q Fb/(2k_{\rm B} T)\right]}.
\label{eq:psim}
\end{align}
Equations (\ref{eq:psis})-(\ref{eq:psis1}) can be generalized to 
\begin{align}
\hat{\psi_{\rm t} }(s) =& 1 - \,_2F_1 \left[1, \alpha, \alpha+1, - \gamma_{\rm rt}(F) /s \right]
\label{eq:psist}\\
\sim& 1-\frac{\pi \alpha}{\sin \pi \alpha} \left( \frac{s}{\gamma_{\rm rt}(F)} \right)^\alpha , 
\label{eq:psist1}
\end{align}
where $\gamma_{\rm r}$ in Eq. (\ref{eq:psit}) is replaced by $\gamma_{\rm rt}(F)$ given by Eq. (\ref{eq:gamma_rt}).

In CTRW theory, the Fourier transform of the probability density for carriers at $x$ starting from $x_{\rm i}$ in free space in the absence of boundaries is given in the Laplace domain as \cite{hughes_95}
\begin{align}
\hat{G}_0 (k,s)&=\int_{-\infty}^\infty dx\, \exp[ik(x-x_{\rm i})] \hat{G}_0 (x,x_{\rm i},s), 
\label{eq:invg0}\\
&=\frac{1-\hat{\psi }_{\rm t} (s) }{s}\frac{1}{1-\hat{\psi}_{\rm t} (s) \lambda(k)},  
\label{eq:invg0s}
\end{align}
where $\lambda(k)$ indicates the structure factor for the random walk; 
for a one-dimensional unbiased random walk, $\lambda (k)=(1/2)\sum_{x=\pm b} \exp(i k x)=\cos (k b)\approx 1-(kb)^2/2$, where the approximation is carried out under the assumption that $kb\ll1$. 
The factor $(1-\hat{\psi}_{\rm t} (s))/s$ is the probability that trapped carriers remain without transitions to new sites up to time $t$, given by 
\begin{align}
\varphi_{\rm t} (t)=\int_t^\infty dt_1\, \psi_{\rm t} (t_1).
\label{eq:remain}
\end{align}
The Laplace transform of $\varphi_{\rm t} (t)$ is $\hat{\varphi}_{\rm t} (s)=(1-\hat{\psi }_{\rm t} (s))/s$.
As shown in Appendix A, the structure factor $\lambda (k)$ under the bias can be expressed as 
\begin{align}
\lambda(k) \approx 1 +iA k-\frac{B}{2} k^2,  
\label{eq:C}
\end{align}  
where $A$ and $B$ are obtained as 
\begin{align}
A&=b\left[\Gamma_p (F)-\Gamma_m(F)\right]=b\tanh[qFb/(2k_{\rm B} T)]\approx qFb^2/(2k_{\rm B} T),
\label{eq:Aapprox}\\
B&=b^2 ,
\label{eq:Bapprox}
\end{align}
using Eqs. (\ref{eq:psip})--(\ref{eq:psim}).  
The Green's function in free space should be translationally invariant, and the Laplace transform of $G_0(x,x_{\rm i},t)$ can be written as
\begin{align}
\hat{G}_0(x,x_{\rm i},s)&= \frac{1-\hat{\psi}_{\rm t}}{s}\frac{1}{2\pi} \int_{-\infty}^\infty dk\, \frac{\exp[-ik(x-x_{\rm i})]}{1-\hat{\psi}_{\rm t} \lambda(k)}
\label{eq:G0_0d}\\
&\approx \frac{1-\hat{\psi}_{\rm t}}{s}\frac{1}{2\pi} \int_{-\infty}^\infty dk\, \frac{\exp[-ik(x-x_{\rm i})]}{1-\hat{\psi}_{\rm t} \left[1+iAk-(B/2)k^2 \right]},
\label{eq:G0}
\end{align}
where $x_{\rm i}$ is the initial position.
As shown in Appendix B, $\hat{G}_0(x,x_{\rm i},s)$ can be expressed as \cite{Weiss_94}
\begin{align}
G_0(x,x_{\rm i},t)=\exp\left[(A/B)(x-x_0)\right]g_0(|x-x_{\rm i}|,t),  
\label{eq:G0_1}
\end{align}
where $\hat{g}_0(|x-x_{\rm i}|,s)$ is given by
\begin{align}
\hat{g}_0(|x-x_{\rm i}|,s)=\frac{1-\hat{\psi}_{\rm t} }{s\sqrt{\hat{\psi}_{\rm t}\left[2 B\left(1- \hat{\psi}_{\rm t}\right)+A^2 \hat{\psi}_{\rm t}
\right]}}
\exp
\left[-\frac{A}{B}
\left(|x-x_{\rm i}|
\sqrt{1+
\frac{2B\left(1- \hat{\psi}_{\rm t}\right)}{A^2\hat{\psi}_{\rm t}
}
}
\right) 
\right] . 
\label{eq:G0_2}
\end{align}
In the numerical evaluation, we use Eqs. (\ref{eq:G0_1}) and (\ref{eq:G0_2}) with Eq. (\ref{eq:psist}).  
 
We now consider the influence of the boundary condition where charge carriers are extracted from the charge-generation layer at $x=L$. 
In the Scher--Montroll theory, the boundary condition at $x=0$ is not imposed because  the reflecting boundary at $x=0$ can be ignored for the carriers moving in the positive direction of the $x$-axis. 
However, the influence of carriers extracted from $x=L$ is considered. 
The effect of carrier extraction at $x=L$ can be taken into account by introducing the probability density for the first passage time to reach $x=L$ starting from $x_{\rm i}$ denoted by $f(L, x_{\rm i}, t)$.
The carrier probability density at $x$ starting from $x_{\rm i}$ at time $t$ is denoted by $G (x,x_{\rm i},t)$ under the influence of the boundary at $x=L$, where carriers are perfectly extracted.  
$G (x,x_{\rm i},t)$ can be expressed using the carrier probability density in free space [$G_0 (x,x_{\rm i},t)$] by subtracting the carrier probability density that would have reached $L$ at some earlier time $t_1$ and then propagated back to $x$: \cite{Scher_75}
\begin{align}
G (x,x_{\rm i},t)=G_0 (x,x_{\rm i},t)-\int_0^t dt_1\, f(L, x_{\rm i}, t) G_0 (x,L,t-t_1). 
\label{eq:SM}
\end{align}
The Laplace transform of $G (x,x_{\rm i},t)$ can be expressed as 
\begin{align}
\hat{G} (x,x_{\rm i},s)=\hat{G}_0 (x,x_{\rm i},s)-\hat{f}(L, x_{\rm i}, s) \hat{G}_0 (x,L,s). 
\label{eq:SMs}
\end{align}
When the carriers are perfectly extracted at $x=L$, we have $G (L,x_{\rm i},t)=0$ and 
\begin{align}
\hat{f}(L, x_{\rm i}, s)=\frac{\hat{G}_0 (L,x_{\rm i},s)}{\hat{G}_0 (0,s)}. 
\label{eq:f}
\end{align}
As shown in Appendix B, $\hat{f}(L, x_{\rm i}, s)$ can be transformed into  \cite{Weiss_94}
\begin{align}
\hat{f}(L, x_{\rm i}, s)=\exp
\left[\frac{A}{B}
\left(L-x_{\rm i}-|L-x_{\rm i}|
\sqrt{1+
\frac{2B\left(1- \hat{\psi}_{\rm t}\right)}{A^2\hat{\psi}_{\rm t}
}
}
\right) 
\right] . 
\label{eq:f_1}
\end{align}
Equation (\ref{eq:SMs}), together with Eq. (\ref{eq:f_1}), constitutes the basis to study the influence of carrier extraction at $x=L$. 

If we denote the initial distribution of injected carriers by $p_{\rm i} (x_{\rm i})$, 
the carrier profile can be calculated from
\begin{align}
P_{\rm r} (x,t)=\int_0^L dx_{\rm i}\, G (x,x_{\rm i},t)p_{\rm i} (x_{\rm i}),
\label{eq;profile}
\end{align}
and we denote the probability density for carriers that survived extraction in the carrier generation layer by 
\begin{align}
S (t)=\int_0^L dx P_{\rm r} (x,t).  
\label{eq:S_1_i}
\end{align}
The transient photocurrent under the static applied voltage can be calculated from (see Appendix C for details) 
 \cite{Nishizawa_06,Philippa_11,Sagues_17}
\begin{align}
J(t)=-\frac{q}{L} \frac{d}{dt} \int_0^L dx \int_0^L dx_{\rm i} (L-x)G (x,x_{\rm i},t) p_{\rm i} (x_{\rm i}), 
\label{eq:constePot}
\end{align}
where $q$ indicates the elementary charge and where we ignored a proportionality constant because we will focus on the time-dependence of $J(t)$.
The transient current is expressed by Eq. (\ref{eq:constePot}) as a relative value because the carrier distribution is used instead of the carrier density.  

Numerical calculations are performed when the initial charge-carrier distribution is expressed by
\begin{align} 
p_{\rm i} (x)=\beta \exp\left( -\beta x
\right)/\left[1-\exp(-\beta L)\right] .
 \label{eq:pi}
 \end{align} 
For a uniform initial carrier distribution and a localized initial carrier distribution at $x=0$, $p_{\rm i} (x)$ can be expressed as 
\begin{align} 
p_{\rm i} (x)=
\begin{cases}
\displaystyle 
\frac{1}{L} \\
\displaystyle 
\delta(x) ,
\end{cases}
 \label{eq:piud}
  \end{align} 
respectively. Equations (\ref{eq;profile})--(\ref{eq:constePot}) can be calculated using the Laplace transformation, where the Laplace transform of $G (x,x_{\rm i},t)$ is obtained from Eq. (\ref{eq:SMs}) using $\hat{f}(L, x_{\rm i}, s)$ given by Eq. (\ref{eq:f_1}). in Eqs. (\ref{eq:SMs})--(\ref{eq:f}), $\hat{G}_0(x,x_{\rm i},s)$ is obtained from Eqs. (\ref{eq:G0_1}) and (\ref{eq:G0_2}) with Eq. (\ref{eq:psist}). For simplicity, we did not impose the additional reflecting boundary condition at $x=0$, which, in principle, can be taken into account as shown in Appendix D. When the initial charge-carrier distribution is given by the superposition of the exponential functions, the analytical expressions in Eqs. (\ref{eq;profile}) --(\ref{eq:constePot}) in the Laplace domain can be obtained using Mathematica. \cite{Mathematica}
The analytical expressions in Eqs. (\ref{eq;profile}) --(\ref{eq:constePot}) can be also obtained in the Laplace domain for the uniform and delta-function initial carrier distributions. 
Equations (\ref{eq;profile}) --(\ref{eq:constePot}) in the time domain are calculated by a numerical inverse Laplace transformation using the Stehfest method. \cite{Stehfest1970_47}
All the numerical results are presented using the dimensionless time unit given by $\gamma_{\rm r} t$; time is normalized by the rate constant associated with hopping.  
The transient currents are shown as relative values, where the time dependence is maintained. 

\section{The Scher--Montroll theory: Localized carrier injection}

First, we consider the transient photocurrent in free space where the initial carriers are generated at $x=0$; we ignore the influence of charge-carrier extraction at $x=L$. 
For simplicity, we also ignore the reflecting boundary condition at $x=0$.
Using Eq. (\ref{eq:invg0}), we obtain the mean of $x$ in free space by  
\begin{align}
\langle \hat{x} (s) \rangle_{\rm f} =\left. -i \frac{\partial \hat{G}_{0} (k,s)}{\partial k} \right|_{k=0} ,
\label{eq:avx}
\end{align}
where $\langle \hat{x}\rangle_{\rm f} $ indicates the Laplace transform of $\langle x(t) \rangle_{\rm f} $ and the subscript "f" indicates the quantity in free space; 
$\langle x(t) \rangle_{\rm f}=\int_{-\infty}^{\infty} dx x G_0(x,0,t)$. 
By substituting Eq. (\ref{eq:invg0s}) together with Eqs. (\ref{eq:psis1}) and (\ref{eq:C}) into Eq. (\ref{eq:avx}), we obtain 
\begin{align}
\langle \hat{x} (s)\rangle_{\rm f} =\frac{\sin \pi \alpha}{\pi \alpha} \frac{A\gamma_{\rm rt}^\alpha}{s^{1+\alpha}}.
\label{eq:x0}
\end{align}
The current in free space denoted by $I_{\rm f}$ should be proportional to the inverse Laplace transform of $q s \langle \hat{x} \rangle_{\rm f}$, which is obtained as
\begin{align}
I_{\rm f} (t) \propto q\frac{\partial }{\partial t} \langle x(t) \rangle_{\rm f} =\frac{\sin \pi \alpha}{\pi \alpha}
\frac{A\gamma_{\rm rt}^\alpha}{\Gamma(\alpha) t^{1-\alpha}} \mbox{ for } t<t_{\rm tr}, 
\label{eq:v0}
\end{align}
where $t_{\rm tr}$ indicates the transit time, which differentiates the initial time scale for free-carrier motion from the later time regime, where the carrier density is reduced by the fraction of carriers passing through the carrier-extracting boundary.
Therefore, the initial current decays by $1/t^{1-\alpha}$ time dependence when the charge-extracting boundary does not influence the current. 
This is the essence of the Scher--Montroll theory with regard to the initial exponent. \cite{Scher_75,Pfister_78}

\subsection{Photocurrent of charge carriers in region between $x=0$ and $x=L$}
Here, we consider the transient photocurrent in free space but also consider that the current should originate from charge carriers between $x=0$ and $x=L$. 
Here, the boundary condition at $x=L$ has not yet been imposed. 
In the subsequent subsection, we show the result of setting the boundary condition that the charge-carrier probability density at $x=L$ is zero to express the influence of charge-carrier extraction at $x=L$. 
For simplicity, we also ignore the reflecting boundary condition at $x=0$ and consider  
\begin{align}
I_0 (t) \propto q\frac{\partial }{\partial t} \int_0^L dx\, x G_0 (x,x_{\rm i},t).
\label{eq:x0def}
\end{align}
In the Scher--Montroll theory, carriers are assumed to be initially located at $x_{\rm i}=0$ and the current is defined using Eq. (\ref{eq:x0def}) rather than Eq. (\ref{eq:constePot}); \cite{Scher_75,Pfister_78} 
the constant applied voltage maintained by charging and discharging of electrodes 
to compensate the internal field change induced by carrier transport 
is taken into account in Eq. (\ref{eq:constePot}), 
while such effect is ignored in Eq. (\ref{eq:x0def}). 
By substituting Eq. (\ref{eq:G0_1}) with Eq. (\ref{eq:G0_2}) into 
$\langle \hat{x} (0,s) \rangle_0=\int_0^L dx\, x \hat{G}_0 (x,0,s)$, 
we obtain 
\begin{align}
\langle \hat{x} (0,s) \rangle_0
&= 
\frac{B \left(1-\hat{\psi}_{\rm t} \right)\left[B
-e^{\frac{AL(1-R)}{B}} 
\left(AL \left(R-1\right)+B\right) \right]
}{s A^3
R \left(R-1\right)^2 
\hat{\psi}_{\rm t}}
\label{eq:x0L}\\
&\approx 
\begin{cases}
\displaystyle 
\frac{B^2 \left(1-\hat{\psi}_{\rm t} \right)}{s A^3
R \left(R-1\right)^2 
\hat{\psi}_{\rm t}} ,
&\mbox{ for } AL\left(R-1\right) > B, 
\\
\displaystyle \frac{B^2 \left(1-\hat{\psi}_{\rm t} \right)\left(1
-e^{\frac{AL(1-R)}{B}} 
\right)
}{s A^3
R \left(R-1\right)^2 
\hat{\psi}_{\rm t}}
&\mbox{ for } AL\left(R-1\right) < B ,
\end{cases} 
\label{eq:x0L_1}
\end{align}
where we define 
\begin{align}
R=\sqrt{1+2 (B/A^2) (1-\hat{\psi}_{\rm t})/\hat{\psi}_{\rm t}}  . 
\label{eq:R}
\end{align}
We note that
\begin{align}
R-1 \approx (B/A^2)(1-\hat{\psi}_{\rm t})/\hat{\psi}_{\rm t}
\approx \frac{B}{A^2}\frac{\pi \alpha}{\sin \pi \alpha} \left( \frac{s}{\gamma_{\rm rt}(F)} \right)^\alpha
\label{eq:Rapprox}
\end{align}
using Eq. (\ref{eq:psist1}) and that $R-1 \propto (s/\gamma_{\rm rt}(F) )^\alpha$ tends to zero as $s \rightarrow 0$. 
The result suggests that the lower term in Eq. (\ref{eq:x0L_1}) is appropriate in the limit of  $s \rightarrow 0$.
More rigorously, we can define $s^*$ satisfying the condition obtained from $AL\left(R-1\right) =B $ as 
\begin{align}
L(\sqrt{1+2 (B/A^2) (1-\hat{\psi}_{\rm t}(s^*))/\hat{\psi}_{\rm t}(s^*)}-1)=B/A.
\label{eq:sstar}
\end{align}
$s^*$ corresponds to the inverse of the transit time given by Eq. (\ref{eq:TOFtra}). Using  Eq. (\ref{eq:x0L}), we obtain
\begin{align}
s\langle \hat{x} (0,s) \rangle_0\approx \begin{cases}
\displaystyle A \frac{\sin \pi \alpha} {\pi \alpha}\frac{\gamma_{\rm rt}^\alpha }{s^{1+\alpha} } +
A
\frac{\pi \alpha}{\sin \pi \alpha} \frac{s^{\alpha-1} }{\gamma_{\rm rt}^\alpha }&\mbox{ for } s>s^*\\
\displaystyle 
\frac{L^2}{2A } 
\frac{\pi \alpha}{\sin \pi \alpha} \frac{s^{\alpha-1} }{\gamma_{\rm rt}^\alpha }
&\mbox{ for } s<s^*. 
\end{cases} 
\label{eq:x0Ls}
\end{align}
Applying the inverse Laplace transformation to Eq. (\ref{eq:x0Ls}), we obtain the transient current for the TOF setting, where the initial carriers are generated at $x_{\rm i}=0$, as 
\begin{align}
I (t) \propto 
\begin{cases}
\displaystyle \frac{\sin \pi \alpha}{\pi \alpha}
\frac{A\gamma_{\rm rt}^\alpha}{\Gamma(\alpha) t^{1-\alpha}}+ \frac{\alpha}{\Gamma(1-\alpha)} 
\left(\frac{\pi \alpha}{\sin \pi \alpha} \right)\frac{A}{\gamma_{\rm rt}^{\alpha}t^{1+\alpha}} &\mbox{ for } t<t_{\rm tr}\\
\displaystyle \frac{\alpha}{\Gamma(1-\alpha)} 
\left(\frac{\pi \alpha}{\sin \pi \alpha} \right)\frac{L^2}{2A}\frac{1}{\gamma_{\rm rt}^{\alpha}t^{1+\alpha}}
&\mbox{ for } t>t_{\rm tr}. 
\end{cases}
\label{eq:uni}
\end{align}
By equating the first terms on the right-hand sides of Eq. (\ref{eq:uni}), 
\begin{align}
\frac{\sin \pi \alpha}{\pi \alpha}
\frac{A\gamma_{\rm rt}^\alpha}{\Gamma(\alpha) t_{\rm tr}^{1-\alpha}}=\frac{\alpha}{\Gamma(1-\alpha)} 
\left(\frac{\pi \alpha}{\sin \pi \alpha} \right)\frac{L^2}{2A}\frac{1}{\gamma_{\rm rt}^{\alpha}t_{\rm tr}^{1+\alpha}},
\nonumber 
\end{align}
where the proportionality constant is the same, we obtain 
\begin{align}
t_{\rm tr}
&\approx \frac{1}{\gamma_{\rm rt}(F)} \left( \frac{\pi \alpha}{\sin \pi \alpha} \right)^{1/\alpha}
\left(\frac{\Gamma(1+\alpha)}{2\Gamma(1-\alpha)}
\right)^{1/(2\alpha)}
\left(\frac{L
}{A(F)}\right)^{1/\alpha}
\propto \frac{\left[(L/b) \coth(qFb/(2k_{\rm B} T))\right]^{1/\alpha}}{\gamma_{\rm r}\cosh[qFb/(2k_{\rm B} T)] }
\label{eq:TOFtra}\\
&\propto (L/b)^{1/\alpha}\left[2k_{\rm B} T/(qFb)\right]^{1/\alpha}/\gamma_{\rm r} \mbox{ for } qFb/(2k_{\rm B} T)<1 ,  
\label{eq:TOFtra1}
\end{align}
where we have used $\Gamma(1+\alpha)=\alpha \Gamma(\alpha)$. \cite{NIST}
The external field dependence of the factor characterizing the difference between the forward release rate and the backward release rate for a hopping transition denoted by $A(F)$ is given by Eq. (\ref{eq:Aapprox}). 
The external field dependence of the total release rate denoted by $\gamma_{\rm rt}(F)$ is given by Eq. (\ref{eq:gamma_rt}). 
According to Eqs. (\ref{eq:Aapprox}) and (\ref{eq:gamma_rt}), $A(F)$ and $\gamma_{\rm rt}(F)$ are proportional to $\tanh[qFb/(2k_{\rm B} T)]$ and $\cosh[qFb/(2k_{\rm B} T)]$, respectively.
The result can be regarded as the generalization of the conventional expression, where  we have $A \propto qFb/(2k_{\rm B} T)$ and the field-independent $\gamma_{\rm rt}$ under a weak electric field.  \cite{Scher_75,Pfister_78} 
The transit time given by Eq. (\ref{eq:TOFtra1}) with $A \propto qFb/(2k_{\rm B} T)$ has been examined experimentally using the TOF method. \cite{Scher_75,Pfister_78,Tiedje_81,Nebel_89} 
In a high-strength electric field ($1 \times 10^6$ to $1 \times 10^7$ V/m$^{-1}$), $A$ in Eq. (\ref{eq:TOFtra}) should be generalized from $A \propto qFb/(2k_{\rm B} T)$, as pointed out experimentally. \cite{MURAYAMA_92}
Under a strong electric field, the field dependence in $\gamma_{\rm rt}(F)$ should also be considered, along with the full-field dependence in $A(F)$, as shown in Eq. (\ref{eq:TOFtra}).  
Equations (\ref{eq:TOFtra})--(\ref{eq:TOFtra1}) also show that, apart from a numerical factor, the inverse of $t_{\rm tr}$ indicates the hopping-release frequency in the absence of an external electric field ($\gamma_{\rm r}$) when $qFb/(k_{\rm B} T)$ and $L/b$ are known in advance. 
If the diffusion is dispersive, where a clear characteristic time scale for hopping transport is absent, $\gamma_{\rm r}$ represents the maximum hopping frequency, $E=0$ in Eq. (\ref{releaser}); the maximum hopping frequency can be estimated from the inverse of the transit time, although the diffusion is dispersive.

\subsection{Influence of charge-carrier extraction at $x=L$}
By multiplying $x$ and integrating both sides of Eq. (\ref{eq:SMs}), we obtain 
\begin{align}
\langle \hat{x} (x_{\rm i},s) \rangle=\langle \hat{x} (x_{\rm i},s) \rangle_0 -\hat{f}(L, x_{\rm i}, s) \langle \hat{x} (L,s) \rangle_0 , 
\label{eq:x_1}
\end{align}
where $\langle \hat{x} (x_{\rm i},s) \rangle$ indicates $\langle \hat{x} (x_{\rm i},s) \rangle=\int_0^L dx\, x \hat{G} (x,x_{\rm i},s)$.

By substituting Eq. (\ref{eq:G0_1}) with Eq. (\ref{eq:G0_2}) into Eq. (\ref{eq:x0def}) and using (\ref{eq:f_1}), we obtain 
\begin{align}
\hat{f}(L, 0, s) \langle \hat{x} (L,s) \rangle_0&=\frac{e^{\frac{-2ALR}{B}} B^2 \left(1-\hat{\psi}_{\rm t} \right)
+e^{\frac{AL \left(1-R\right)}{B}} \left[AL \left(R+1\right)-B\right]
}{s A^2 R
\left(R+1\right)^2 \hat{\psi}_{\rm t} } 
\label{eq:fxL1}\\
&\approx -\frac{B(2AL-B)}{4A^3 s} \frac{\pi \alpha}{\sin \pi \alpha} \frac{s^{\alpha} }{\gamma_{\rm rt}^\alpha } ,
\label{eq:fxL2}
\end{align}
where $R$ is given by Eq. (\ref{eq:R}). 

By substituting Eqs. (\ref{eq:x0Ls}) and (\ref{eq:fxL2}) into Eq. (\ref{eq:x_1}) and using $\langle \hat{v} \rangle=s \langle \hat{x} \rangle$, we obtain the Laplace transform of the current as 
\begin{align}
\hat{I} (s)\propto  \langle \hat{v}(0,s) \rangle \approx 
\begin{cases}
\displaystyle A \frac{\sin \pi \alpha} {\pi \alpha}\frac{\gamma_{\rm rt}^\alpha }{s^{1+\alpha} }+
A
\frac{\pi \alpha}{\sin \pi \alpha} \frac{s^{\alpha-1} }{\gamma_{\rm rt}^\alpha } &\mbox{ for } s>s^*\\
\displaystyle
\frac{L^2}{2A } 
\frac{\pi \alpha}{\sin \pi \alpha} \frac{s^{\alpha-1} }{\gamma_{\rm rt}^\alpha }
&\mbox{ for } s<s^*, 
\end{cases}
\label{eq:qvLLaplace}
\end{align}
where we introduce $2AL>B$.
Equation (\ref{eq:qvLLaplace}) is equal to Eq. (\ref{eq:uni}) obtained without adjusting the boundary condition at $x=L$, where the proportionality constant is the same.
The result indicates that $\langle \hat{v} \rangle=s \langle \hat{x} \rangle\approx s \langle \hat{x} (0,s)\rangle_0$. 
Therefore, the boundary condition at $x=L$ is required to precisely determine the transient photocurrent, whereas the overall decay can be obtained without imposing the boundary condition at $x=L$. 
This fact also holds when the transient current is nondispersive and provides the theoretical basis of the analytical expression for the nondispersive photocurrent decay derived without taking into account the boundary condition at $x=L$. \cite{Hirao_95,Nishizawa_06}

\section{In-depth carrier injection}

\begin{figure}[h]
\begin{center}
\includegraphics[width=10cm]{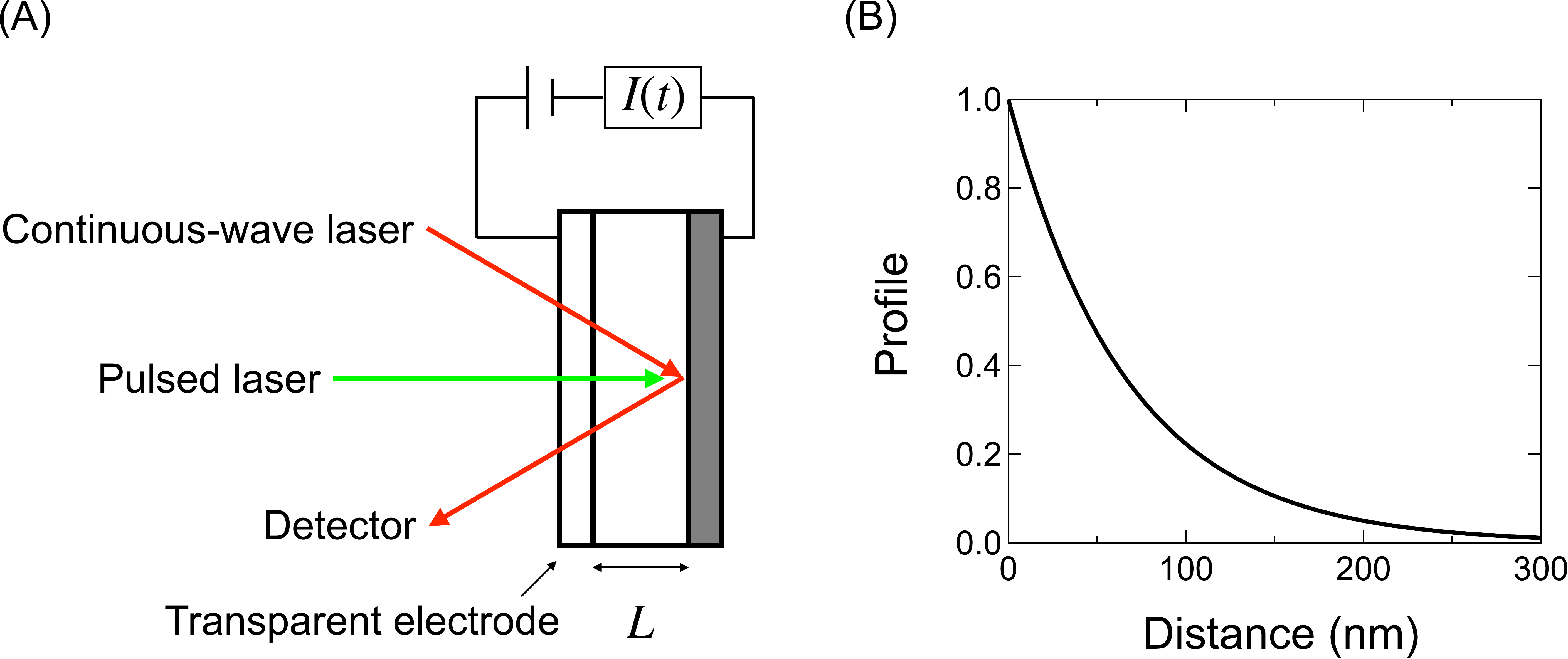}
\end{center}
\caption{(Color online) (A) Schematic of transient photocurrent and density measurements for film conductor with thickness 
$L=300$ nm. 
(B) Profile $p_{\rm i} (x)/p_{\rm i} (0)$ for charge carriers generated by pulsed light. 
Equation (\ref{eq:pi}) is used, where 
$\beta=0.016$ nm$^{-1}$. 
}
\label{fig:1}
\end{figure}

Fig. \ref{fig:1} (A) shows a schematic of the transient photocurrent and carrier density measurements. 
The pulsed light irradiates the film conductor through the transparent electrode. 
The photocurrent response and the optical absorption, which is a measure of the charge-carrier density, are measured simultaneously. 
The latter is measured by detecting optical absorption using weak continuous-wave light.   
The initial distribution of injected carriers is shown in Fig. \ref{fig:1} (B). 
The values of $L=300$ nm and $\beta=0.016$ nm$^{-1}$ used are typical values for organic solar cells. \cite{Nakami_17}
If the value of $L$ is decreased, we need to take into account the reflection from the back electrode. 
Later, we show the results of uniform initial carrier injection. 
The initial carrier distribution for the thickness of $L=45$ nm or less deviates at most 20\% from 
the uniform carrier distribution 
by considering the reflection from the back electrode and $\beta=0.016$ nm$^{-1}$.
\begin{figure}[h]
\begin{center}
\includegraphics[width=6.5cm]{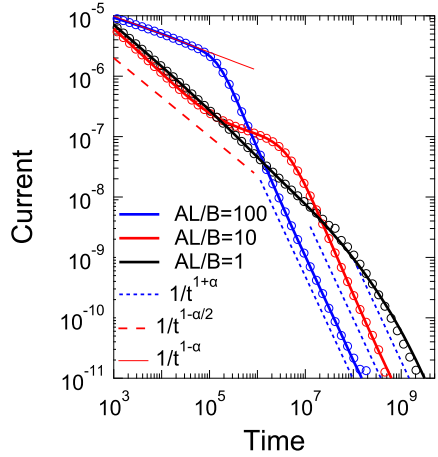}
\end{center}
\caption{(Color online) Transient current (relative) 
calculated for $b=0.5$ nm, $\beta=0.016$ nm$^{-1}$, and $L=300$ nm.
The dimensionless time unit given by $\gamma_{\rm r} t$ is used, and $\alpha=0.7$. 
The thick black line, thick red line, and thick blue line indicate $AL/B\approx qFL/(2k_{\rm B} T)=1$, $10$, and $100$, respectively. 
The thick lines are calculated using Eq. (\ref{eq:constePot}) with Eqs. (\ref{eq:SMs}), (\ref{eq:G0_1}), and (\ref{eq:f_1}).
The open circles with the corresponding color indicate the results obtained using Eq. (\ref{eq:G0_1}), where the boundary conditions are not considered.
The blue dashed line, red dashed line, and red thin line indicate $1/t^{1+\alpha}$[the second line of Eq. (\ref{eq:uni})], $1/t^{1-\alpha/2}$ [Eq. (\ref{eq:Idiff_1})], and $1/t^{1-\alpha}$ [the first line of Eq. (\ref{eq:uni})], respectively. 
The initial charge density profile is shown in Fig. \ref{fig:1} (B). 
}
\label{fig:2}
\end{figure}

The numerical results for the transient photocurrent are presented in Fig. \ref{fig:2}. 
The transition from a $1/t^{1-\alpha}$ time dependence to a $1/t^{1+\alpha}$ time dependence, as predicted by the Scher--Montroll theory, occurs when the external field strength is high ($AL/B=100$).
When the field strength is weak ($AL/B=1$), we obtain the transition from a $1/t^{1-\alpha/2}$ time dependence to a $1/t^{1+\alpha}$ time dependence.
At an intermediate field strength ($AL/B=10$), we observe a series of transitions: $1/t^{1-\alpha/2}$, $1/t^{1-\alpha}$, and $1/t^{1+\alpha}$ time dependence. 

For in-depth carrier injection, carriers can migrate to the extracting boundary by diffusion rather than by drift. 
Under normal diffusion with the diffusion constant denoted by $D$, the mean square displacement in free space denoted by $\sqrt{\langle x(t)^2 \rangle_{\rm f}}$ obeys $\sqrt{\langle x(t)^2 \rangle_{\rm f}}=\sqrt{2Dt}$.
The spreading of the diffusive carriers might be important in the short-time regime because $\sqrt{2Dt}$ is a convex function of time and the drift of the carrier position under the external electric field can be given using a constant velocity $v$ as $vt$, which is not a convex function of time.  
In this case, the current can be obtained from $I_{\rm diff}\propto\partial \sqrt{\langle x(t)^2 \rangle_{\rm f}}/(\partial t)=(\sqrt{D/2}) /t^{1/2}$, where the subscript "diff" indicates the diffusive current. 
The diffusive current can be observed in the early-time regime before the drift current induced by the external electric field dominates. 
In the case of sub-diffusion, the above result can be generalized by noting that
$\langle x(t)^2 \rangle_{\rm f} =2D_{\alpha} t^\alpha$ and
\begin{align}
I_{\rm diff}\propto\frac{\partial}{\partial t} \sqrt{\langle x(t)^2 \rangle_{\rm f}}=
\alpha \left(\sqrt{D_\alpha/2 }\right)/t^{1-\alpha/2},
\label{eq:Idiff}
\end{align}
where $D_\alpha$ indicates the diffusion coefficient for carriers under sub-diffusion motion. 
Although the possibility of the transient current decay given by Eq. (\ref{eq:Idiff}) has been pointed out previously, \cite{Sagues_17} 
the condition under which the transient current decay in Eq. (\ref{eq:Idiff}) is obtained has not yet been studied in terms of in-depth carrier injection. 
By assuming uniform initial carrier injection, $p_{\rm i} (x)=1/L$, the more precise calculation shown in Appendix E 
yields 
\begin{align}
I (t) \propto 
\displaystyle \frac{LA}{8\Gamma(\alpha/2)} \sqrt{\frac{2}{B} \frac{\sin \pi \alpha}{\pi \alpha}}
\frac{\gamma_{\rm rt}^{\alpha/2}}{ t^{1-\alpha/2}}  .
\label{eq:Idiff_1}
\end{align}
Therefore, when the external electric field is weak, the initial current decays with a $1/t^{1-\alpha/2}$ time dependence under the in-depth carrier injection. 
When the electric field is increased, the carrier transport is more influenced by the external electric field and the time dependence is given by  $1/t^{1-\alpha}$ in Eq. (\ref{eq:v0}). 
Nevertheless, the initial decay can still be expressed by a $1/t^{1-\alpha/2}$ time dependence of diffusive currents. 

By equating the asymptotic solution of Eq. (\ref{eq:uni}) with Eq. (\ref{eq:Idiff_1}), where the proportionality coefficient is the same, we obtain the transit time for this case as 
\begin{align}
t_{\rm trd}
&\approx \frac{1}{\gamma_{\rm rt}(F)} \left( \frac{\pi \alpha}{\sin \pi \alpha} \right)
\left(\frac{\alpha \Gamma(\alpha/2)}{\Gamma(1-\alpha)}
\right)^{2\alpha/3}
\left(\sqrt{\frac{B(F)}{2}}\frac{4L
}{A(F)^2}\right)^{2\alpha/3}
\propto \frac{\left[(L/b) \cosh(qFb/(2k_{\rm B} T))\right]^{2\alpha/3}}{\gamma_{\rm r}\left[\sinh(qFb/(2k_{\rm B} T))\right]^{4\alpha/3} }
\label{eq:TOFtrad}\\
&\propto (L/b)^{2\alpha/3}\left[2k_{\rm B} T/(qFb)\right]^{4\alpha/3}/\gamma_{\rm r} \mbox{ for } qFb/(2k_{\rm B} T)<1 . 
\label{eq:TOFtrad1}
\end{align}
The maximum hopping frequency, $\gamma_{\rm r}$, can be estimated from the inverse of the transit time when both $L/b$ and $qFb/(k_{\rm B} T)$ are known. 
The field dependence differs from that in Eqs. (\ref{eq:TOFtra})--(\ref{eq:TOFtra1}) derived for the drift-driven current. 

Figure \ref{fig:2} includes the results obtained using Eq. (\ref{eq:G0_1}), as represented by circles, where the charge-extracting boundary condition is not considered: 
in Eq. (\ref{eq:constePot}), $G (x,x_{\rm i},t)$ is substituted by $G_0 (x,x_{\rm i},t)$ given by Eq. (\ref{eq:G0_1}).
Overall photocurrent decays are reproduced without taking into account the charge-extracting boundary condition at $x=L$. 
Better agreement with the results under the boundary condition at $x=L$ is achieved with increasing external field strength.
As the field strength increases, the mean velocity of charge carriers toward the extracting boundary at $x=L$ increases and the fraction of charge carriers moving toward $x=0$ decreases.
Even without the extracting boundary condition at $x=L$ being imposed, the fraction of charge carriers returning to the region between $0$ and $L$ after passing through the boundary at $L$ decreases with increasing field strength. 
Therefore, the charge-extracting boundary condition at $x=L$ becomes immaterial when the field strength is increased.

\begin{figure}[h]
\begin{center}
\includegraphics[width=6.0cm]{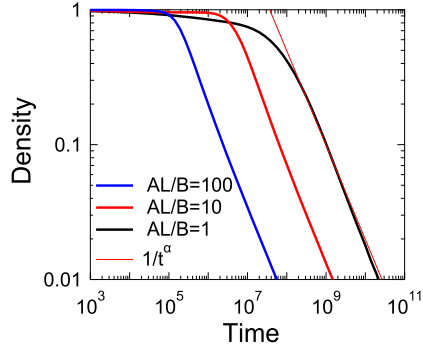}
\end{center}
\caption{(Color online) Probability density for carriers that survive inside the film, as calculated using Eq. (\ref{eq:S_1_i})
for  $b=0.5$ nm, $\beta=0.016$ nm$^{-1}$, and $L=300$ nm.
The dimensionless time unit given by $\gamma_{\rm r} t$ is used, and $\alpha=0.7$.  
The initial charge-density distribution is shown in Fig. \ref{fig:1} (B). 
The black, red, and blue thick solid lines from top to bottom represent $AL/B=100$, $10$, $1$, respectively. 
The red thin line indicates a $1/t^\alpha$ time dependence. 
}
\label{fig:3}
\end{figure}

Figure \ref{fig:3} shows the results for the probability density of carriers that survived inside the film, which corresponds to optical absorption. 
Compared with the transient photocurrents, the transient absorption decay is simpler.
The decay is slow in the initial time regime, where charge-carrier extraction is still limited. 
Substantial decay is observed after the time corresponding to the transit time in the photocurrent. 
The long-time asymptotic decay can be approximated by  $1/t^\alpha$.

\begin{figure}[h]
\begin{center}
\includegraphics[width=12cm]{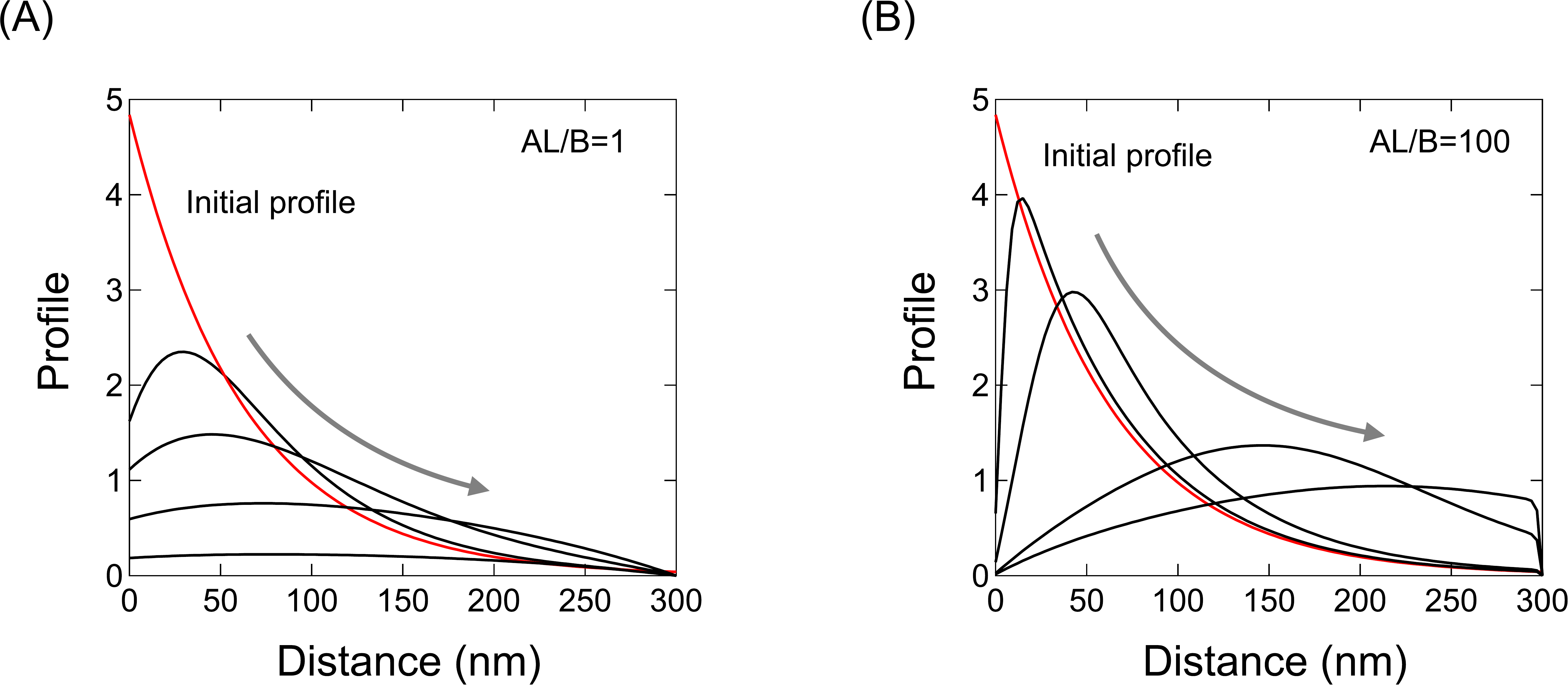}
\end{center}
\caption{(Color online) Density profile in film calculated from Eq. (\ref{eq;profile}) 
for $b=0.5$ nm, $\beta=0.016$ nm$^{-1}$, and $L=300$ nm.
The dimensionless time unit given by $\gamma_{\rm r} t$ is used, and $\alpha=0.7$.
The initial charge density distribution (thin red line) is given by Eq. (\ref{eq:pi}), where 
$\beta=0.016$ nm$^{-1}$, 
as in Fig. \ref{fig:1} (B).   
(A) $AL/B\approx qFL/(2k_{\rm B} T)=1$. 
In (A), the black lines indicate the density profile at dimensionless times of 
$5\times 10^5$, $5\times 10^6$, $5\times 10^7$, and $5\times 10^8$ 
from left to right. 
(B) $AL/B=100$.
In (B), the black lines indicate the density profile at the dimensionless times of 
$1\times 10^3$, $1\times 10^4$, $1\times 10^5$, and $2\times 10^5$ 
from left to right. 
}
\label{fig:4}
\end{figure}

Figure \ref{fig:4} shows the carrier-density profiles for 
$\beta=0.016$ nm$^{-1}$
when $AL/B=1$ and $AL/B=100$. 
In Fig. \ref{fig:4}, the carrier density at $x=L$ is zero, representing the carrier-extracting boundary condition. 
We found that the density profiles are influenced by the extracting boundary condition only in the vicinity of the boundary at $x=L$.
The carrier density profiles decay rapidly to zero in the vicinity of the boundary at $x=L$. 
The steeper decrease of the carrier density around the extracting boundary at $x=L$ is obtained when the field strength is increased from $AL/B\approx qFL/(2k_{\rm B} T)=1$ to $AL/B=100$.
With an increase in the field strength, the influence of the extracting boundary condition on the density profiles is limited to the narrower region.
The initial $1/t^{1-\alpha/2}$ time dependence of the transient current is obtained for $AL/B=1$, and the initial $1/t^{1-\alpha}$ time dependence of the transient current is obtained for $AL/B=100$. 
The carrier-density profile in Fig. \ref{fig:4} (B) [$AL/B=100$] maintains its shape until reaching close to the absorbing boundary at $x=L$, 
which leads to the transition of the transient current from $1/t^{1-\alpha}$ to $1/t^{1+\alpha}$. 
On the other hand, the carrier-density profile in Fig. \ref{fig:4} (A) [$AL/B=1$]  
decays significantly before its peak reaches the absorbing boundary. 
Under the weak electric field [$AL/B\approx qFL/(2k_{\rm B} T)=1$], 
a large part of charge carrier distribution decays by diffusional escape even before the transit-time, 
which leads to the less pronounced transition of the transient current from $1/t^{1-\alpha/2}$ to $1/t^{1+\alpha}$.

\begin{figure}[h]
\begin{center}
\includegraphics[width=6.5cm]{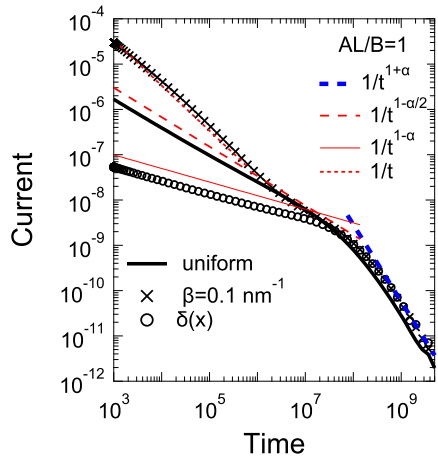}
\end{center}
\caption{(Color online) Transient current (relative) calculated for $AL/B\approx qFL/(2k_{\rm B} T)=1$ 
for $b=0.5$ nm and $L=300$ nm.
The dimensionless time unit given by $\gamma_{\rm r} t$ is used, and $\alpha=0.7$. 
The thick  blackline indicates the uniform initial distribution, whereas the open circles indicate the initial distribution localized at $x=0$, $p_{\rm i} (x)=\delta(x)$. 
The crosses indicate the case where 
$\beta=0.1 $ nm$^{-1}$ in Eq. (\ref{eq:pi}). 
The symbols and the thick line are calculated using Eq. (\ref{eq:constePot}) with Eqs. (\ref{eq:SMs}), (\ref{eq:G0_1}), and (\ref{eq:f_1}).
The blue thick dashed line, red long dashed line, and red thin line are proportional to $1/t^{1+\alpha}$ [the second line of Eq. (\ref{eq:uni})], $1/t^{1-\alpha/2}$ [Eq. (\ref{eq:Idiff_1})], and $1/t^{1-\alpha}$ [the first line of Eq. (\ref{eq:uni})], respectively. 
The red short dashed line indicates $1/t$-time dependence.
}
\label{fig:5}
\end{figure}

Compared with the transient carrier-density decay, the transient currents show rich kinetics; however, the simple transition from $1/t^{1-\alpha}$ to $1/t^{1+\alpha}$ has been predicted in the conventional Scher--Montroll theory.  \cite{Scher_75,Pfister_78}
We here investigate the transition of the transient current from $1/t^{1-\alpha/2}$ to $1/t^{1+\alpha}$ (Fig. \ref{fig:2}) under a weak field in greater detail.
As an extreme case, we consider a uniform density of injected carriers. 
The transient currents obtained for this case are essentially the same as those shown in Fig. \ref{fig:2}.
When $AL/B>1$, the transition from $1/t^{1-\alpha}$ to $1/t^{1-\alpha}$ is obtained for the uniform initial charge density (results not shown).
However, as shown in Fig. \ref{fig:5}, when $AL/B=1$, the initial decay obeys a $1/t^{1-\alpha/2}$ time dependence, as in Fig. \ref{fig:2}.
Figure \ref{fig:5} presents the result when the carriers are generated only at $x=0$ by pulsed light. 
The initial decay for this case clearly shows the $1/t^{1-\alpha}$ time dependence of the conventional Scher--Montroll theory.
We calculate the photocurrent decay by varying $\beta$ in the initial distribution given by Eq. (\ref{eq:pi}); the parameter $\beta$ indicates the inverse of the characteristic length for the initial charge density. 
The results are shown in Fig. \ref{fig:5}. 
Notably, a $1/t^{1-\alpha/2}$ time dependence is obtained even when 
$\beta=0.1$ nm$^{-1}$ 
for $AL/B=1$. 
The condition $AL/B\approx qFL/(2k_{\rm B} T)=1$ corresponds to an electric field strength of 
$1.5 \times 10^5$ V m$^{-1}$ when $L=300$ nm  
and 
$1 \times 10^6$ V m$^{-1}$ when $L=45$ nm. 
Given the field strength and the range of the depth of the carrier distribution shown in Fig. \ref{fig:5}, the transition from a $1/t^{1-\alpha/2}$ time dependence to a $1/t^{1+\alpha}$ time dependence might be observable. 
If the initial decay and the long-time power-law decay are expressed as $1/t^{a_1}$ and $1/t^{a_2}$, respectively, such a transition implies that $a_1+a_2=2+\alpha/2$ rather than  $a_1+a_2=2$ obtained for the classical Scher--Montroll theory, where $a_1=1-\alpha$. 
The initial decay for the case of $\beta=0.1$ nm$^{-1}$ follows $1/t$-time dependence before transition to 
$1/t^{1-\alpha/2}$ time-dependence.  
However, $1/t$-time dependence is  phenomenological because the decay is slightly curved in the log-log plot. 
If the transient current is calculated by Eq. (\ref{eq:x0def}) instead of Eq. (\ref{eq:constePot}), 
the initial decay following $1/t$-time dependence disappears. 
In Eq. (\ref{eq:x0def}), 
charging and discharging of electrodes to compensate the internal field change associated with charge transport is ignored; 
the initial decay following $1/t$-time dependence might originate from charging and discharging effects. 

Although the exponent of the initial power-law decay differs for the carriers generated at $x=0$, the asymptotic decay shows a $1/t^{1+\alpha}$ time dependence for all cases. 
When $\alpha$ in $a_2$ deviates from the relation given by $\alpha=k_{\rm B} T/E_0$, 
the deviation from $a_1+a_2=2$ can be attributed to several factors such as 
the density of states being different from the exponential form given by Eq. (\ref{traped}) and the other model of thermal activation 
but not to the initial distribution of charge carriers.  \cite{Marshall_83,Vanderhaghen_88,MurayamaMori_92,SETO_98} 
The results support the analysis of the transient photocurrent using the time regime later than the transit time to probe the density of states. \cite{Seynhaeve_89,Street_11}
If $\alpha$ in $a_2$ satisfies the relation $\alpha=k_{\rm B} T/E_0<1$, 
one reason for the deviation from $a_1+a_2=2$ 
could be the spatial extent of the initial carriers penetrating the carrier conduction layer.  

\begin{figure}[h]
\begin{center}
\includegraphics[width=6.5cm]{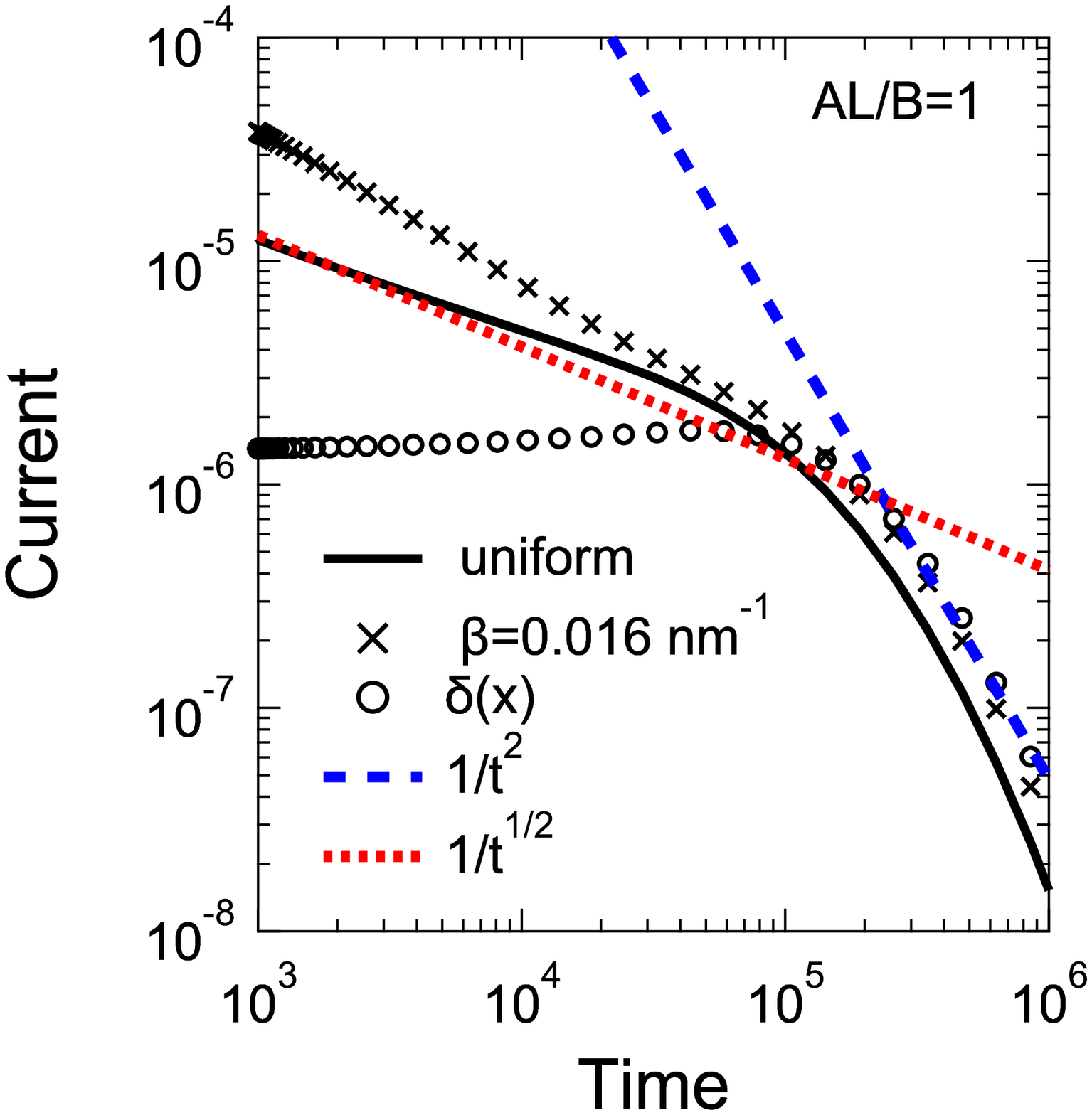}
\end{center}
\caption{(Color online) Transient current (relative) for normal random walk calculated for $AL/B\approx qFL/(2k_{\rm B} T)=1$ 
for $b=0.5$ nm and $L=300$ nm. 
The dimensionless time unit given by $\gamma_{\rm r} t$ is used. 
The thick  black line indicates the uniform initial distribution, whereas the open circles indicate the initial distribution localized at $x=0$, $p_{\rm i} (x)=\delta(x)$. 
The crosses indicate the case where 
$\beta=0.016 $ nm$^{-1}$ in Eq. (\ref{eq:pi}). 
The symbols and the thick line are calculated using Eq. (\ref{eq:constePot}) with Eqs. (\ref{eq:SMs}), (\ref{eq:G0_1}), (\ref{eq:f_1}) and $\hat{\psi_{\rm t} }(s) =\gamma_{\rm rt}/\left(s+\gamma_{\rm rt}\right)$.
The blue long dashed line and red short dashed line are proportional to $1/t^{2}$ and $1/t^{1/2}$, respectively. 
}
\label{fig:6}
\end{figure}
Before closing this section, we present the results of normal random walk for $AL/B\approx qFL/(2k_{\rm B} T)=1$ in Fig. \ref{fig:6}. 
We use  
$\hat{\psi_{\rm t} }(s) = 1/\left[1+(s/\gamma_{\rm rt}) \right]$ instead of Eq. (\ref{eq:psist}). 
In the limit of $s/\gamma_{\rm rt}<1$, we obtain $\hat{\psi_{\rm t} }(s) \approx 1-(s/\gamma_{\rm rt})$, 
which indicates $\alpha=1$. 
When the carriers are generated only at $x=0$ by pulsed light, 
the transient current shows a transition from a plateau regime to $1/t^2$-time dependent regime corresponding to $1/t^{1-\alpha}$-$1/t^{1+\alpha}$ transition, 
followed by an abruptly decay to zero. 
For a uniform density of injected carriers, 
we obtain a transition from $1/t^{1/2}$-time dependence to $1/t^2$-time dependence corresponding to $1/t^{1-\alpha/2}$-$1/t^{1+\alpha}$ transition. 
The transition from $1/t^{1/2}$-time dependence to $1/t^2$-time dependence is also obtained when $\beta=0.016 $ nm$^{-1}$ in Eq. (\ref{eq:pi}). 
We confirm that the transition from plateau regime to $1/t^2$-time dependent regime is obtained for a uniform density of injected carriers when $AL/B\approx qFL/(2k_{\rm B} T)=10$ (results not shown).

\section{Discussion}
\label{sec:Discussion}
We have studied the transient photocurrent and carrier-density decay using the CTRW model.
The model enables us to derive the expressions for the dispersive mobility coefficient and the dispersive diffusion coefficient using the waiting-time distribution for hopping transitions. 
We note that the mobility defined by $\mu(t)=v(t)/(qF)$ is time dependent [$\mu(t) \propto 1/t^{1-\alpha}$] under dispersive kinetics. \cite{Scher_75}
We introduce the mobility coefficient defined by $\mu_\alpha=v(t)\left[qF/(\Gamma(\alpha) t^{1-\alpha}) \right]^{-1}$ to eliminate the $1/t^{1-\alpha}$ time dependence. 
Similarly, in the absence of an external field, we have $\langle x^2 \rangle = 2 D_\alpha t^\alpha$; in addition, $D_\alpha$ indicates the dispersive diffusion coefficient, where $\langle \cdots \rangle$ represents the ensemble average. 

Before studying the relation between the mobility coefficient and the diffusion coefficient, we derive these coefficients from the CTRW model. 
Using the Fourier transformation of Eq. (\ref{eq:G0}), we obtain 
\begin{align}
\hat{G}_0(k,s)
&= \frac{1}{s+[s\hat{\psi}_{\rm t}/(1-\hat{\psi}_{\rm t})]\left[iAk-(B/2)k^2 \right]}, 
\label{eq:aps_2}
\end{align}
where $A$ and $B$ can be expressed as 
\begin{align}
A&=b\tanh[qFb/(2k_{\rm B} T)]\approx qFb^2/(2k_{\rm B} T),
\label{eq:aps_A}\\
B&= b^2 .
\label{eq:aps_B}
\end{align}
By applying the Fourier transformation, we can express for any function $F(x)$  
\begin{align}
&\int_{-\infty}^\infty dk\, \exp(ikx)\left[-\frac{d}{dx} v F(x)+D\frac{d^2}{dx^2} F(x)\right] 
\label{eq:aps_3_1}\\
&=ivk \hat{F}(k)-Dk^2 \hat{F}(k)
\label{eq:aps_3_2}
\end{align}
in the Fourier space.

We first consider the case of normal diffusion. 
Equation (\ref{eq:aps_3_2}) indicates that $[s\hat{\psi}_{\rm t}A/(1-\hat{\psi}_{\rm t})]/F$ and $s\hat{\psi}_{\rm t}B/[2(1-\hat{\psi}_{\rm t})]$ in Eq. (\ref{eq:aps_2}) can be interpreted as the drift velocity and the diffusion constant, where the total hopping frequency is given by $\gamma_r \cosh\left[ q Fb/(2k_{\rm B} T)\right]$ [Eq. (\ref{eq:gamma_rt})]. 
We have $s\hat{\psi}_{\rm t}/(1-\hat{\psi}_{\rm t})=\gamma_{\rm r} \cosh\left[ q Fb/(2k_{\rm B} T)\right]$, and the corresponding drift-diffusion equation can be written as
\begin{align}
\frac{d}{dt} G_{\rm d}(x,t)&=D\left[- \frac{d}{dx} \frac{2\tanh[qFb/(2k_{\rm B} T)]}{b}G_{\rm d}(x,t) +\frac{d^2}{dx^2}G_{\rm d}(x,t) \right],
\label{eq:aps_4_1}\\
\frac{d}{dt} G_{\rm d}(x,t)&\approx D\left[- \frac{d}{dx} \frac{qF}{k_{\rm B} T}G_{\rm d}(x,t) +\frac{d^2}{dx^2}G_{\rm d}(x,t) \right],
\label{eq:aps_4_2}
\end{align}
where the limit of $b \rightarrow 0$ is taken to obtain the last equation, and $G_{\rm d}(x,t)$ just represents the solution of Eq. (\ref{eq:aps_4_1})
in one--dimensional infinite space, 
 which is normalizable, ($\int dx\, G_{\rm d}(x,t)=1$). 
By comparing Eq. (\ref{eq:aps_3_1}) with Eq. (\ref{eq:aps_4_1}). 
the diffusion constant can then be obtained as \cite{Derrida_83,Pautmeier_91}
\begin{align}
D(F)=\gamma_r \cosh\left[ q Fb/(2k_{\rm B} T)\right]b^2/2
\label{eq:DF}
\end{align} 
and the drift velocity as $v(F)=D(2A/b^2)=2D(F) \tanh[qFb/(2k_{\rm B} T)]/b$, where we have $v(F)=d\langle x(t) \rangle/dt$. 
Therefore, we obtain the fluctuation--dissipation relation between the diffusion and drift: 
\begin{align}
\frac{v(F)}{D(F)}=2\tanh[qFb/(2k_{\rm B} T)]/b.
\label{eq:neq}
\end{align}
Equation (\ref{eq:neq}) has been derived previously for a one-dimensional lattice random walk. \cite{Derrida_83,Aslangul_89,Hurowitz_14}
Using Eq. (\ref{eq:aps_4_1}), we obtain
\begin{align}
\frac{d}{dt} \langle \delta x^2(t) \rangle=2D(F) ,
\label{eq:apdeltax}
\end{align}
where $\delta x(t)=x(t)-\langle x(t)\rangle$. 
A measure of the thermodynamic driving force can be given by, \cite{Pautmeier_91}
\begin{align}
\frac{\nabla S}{2k_{\rm b}}= v/\frac{d}{dt} \langle \delta x^2(t) \rangle=\frac{v}{2D(F)}=\tanh[qFb/(2k_{\rm B} T)]/b , 
\label{eq:neq1}
\end{align}
where the entropy gradient is denoted by $\nabla S$ [See also Eq. (\ref{eq:aps_6_1})]. 
We will show that the entropy gradient is independent of $E_0$, 
which indicates the characteristic energy for an exponential density of states, in the CTRW model. 
By defining the mobility as $\mu=v/(qF)$, we find $\mu(F)=2D(F) \tanh[qFb/(2k_{\rm B} T)]/(qFb)$, which can be transformed into \cite{BAGLEY_70,Funabashi_76,Derrida_83,Pautmeier_91}
\begin{align}
\mu(F)=\gamma_r b \sinh[qFb/(2k_{\rm B} T)]/(qF), 
\label{eq:gmu}
\end{align}
or the form of the generalized Einstein relation, 
\begin{align}
D(F)=\mu(F) (q Fb/2)\coth[qFb/(2k_{\rm B} T)].
\label{eq:gEinstein}
\end{align}
In the limit of $F \rightarrow 0$, the conventional Einstein relation recovers 
\begin{align}
D=\mu k_{\rm B} T.
\label{eq:Einstein}
\end{align}

We then consider the case of dispersive diffusion. 
For this case, Eqs. (\ref{eq:aps_4_1})--(\ref{eq:aps_4_2}) can be generalized as
\begin{align}
\frac{d}{dt} G_0(x,t)&=\frac{d}{dt} \int_0^t dt_1 Q(t-t_1)D_\alpha \left[- \frac{d}{dx}\frac{2\tanh[qFb/(2k_{\rm B} T)]}{b}G_0(x,t_1) +\frac{d^2}{dx^2}G_0(x,t_1) \right],
\label{eq:aps_5_1}\\
&\approx\frac{d}{dt} \int_0^t dt_1 Q(t-t_1)D_\alpha \left[- \frac{d}{dx} \frac{qF}{k_{\rm B} T}G_0(x,t_1) +\frac{d^2}{dx^2}G_0(x,t_1) \right],
\label{eq:aps_5_2}
\end{align}
where the Laplace transform of $Q(t)$ is given by ${\cal L}(Q(t))=\hat{\psi}_{\rm t}/(1-\hat{\psi}_{\rm t})$, the generalized diffusion coefficient is given by  
\begin{align}
D_\alpha(F)=\frac{\sin (\pi \alpha)}{\pi \alpha} \gamma_{\rm rt}(F)^\alpha b^2/2
=\frac{\sin (\pi \alpha)}{\pi \alpha}  \left(\gamma_{\rm r}\cosh\left[ q Fb/(2k_{\rm B} T)\right] \right)^\alpha b^2/2,
\label{eq:Dalpha}
\end{align}
and the limit of $b \rightarrow 0$ is taken to obtain the last equation.
Multiplying both sides of Eq. (\ref{eq:aps_5_1}) by $x$ and integrating with respect to $x$, we obtain the average drift velocity as 
\begin{align}
v(t)&=\frac{D_\alpha}{\Gamma(\alpha) t^{1-\alpha}} \frac{2\tanh[qFb/(2k_{\rm B} T)]}{b}, 
\label{eq:aps_mu3}
\end{align}
using the asymptotic form of $Q(t)$ as \cite{BALAKRISHNAN_85,Kenkre_73,Seki_03_2}
\begin{align}
\frac{d}{dt} \int_0^t dt_1 Q(t-t_1)\simeq \frac{d}{d t} \int_0^t dt_1\, \frac{1}{(t-t_1)^{1-\alpha}}=\frac{d}{d t} \int_0^t d\tau\, \frac{1}{\tau^{1-\alpha}}=\frac{1}{t^{1-\alpha}}.
\label{eq:fracdiffop}
\end{align}
Then, we recover Eq. (\ref{eq:neq1}), 
\begin{align}
\frac{\nabla S}{2k_{\rm b}}=v(t)/\frac{d}{dt} \langle \delta x^2(t) \rangle&=\tanh[qFb/(2k_{\rm B} T)]/b 
\label{eq:neq2}\\
&\approx 
\begin{cases}
qF/(2k_{\rm B} T)  &\mbox{ for } qFb/(2k_{\rm B} T)<1\\
\displaystyle
1/b
&\mbox{ for } qFb/(2k_{\rm B} T)>1.
\end{cases}
\label{eq:neq2_1}
\end{align}
The entropy gradient can be understood by noticing that 
the equilibrium distribution is given by $\exp(S/k_{\rm B})$ and 
Eq.  (\ref{eq:aps_5_1}) can be expressed as \cite{Onsager_53,Hashitsume_52,Kubo_73}
\begin{align}
\frac{d}{dt} G_0(x,t)&=\frac{d}{dt} \int_0^t dt_1 Q(t-t_1)\frac{d}{dx} \frac{D_\alpha}{k_{\rm B}} \left[- (\nabla S)G_0(x,t_1) +k_{\rm B}\frac{d}{dx}G_0(x,t_1) \right]. 
\label{eq:aps_6_1}
\end{align}
Multiplying both sides of Eq. (\ref{eq:aps_6_1}) by $x$ and integrating with respect to $x$, we obtain the average drift velocity as 
\begin{align}
v(t)&=\frac{d}{dt} \int_0^t dt_1 Q(t-t_1) \frac{D_\alpha}{k_{\rm B}} \nabla S .
\label{eq:aps_6_v}
\end{align}
Equation (\ref{eq:aps_6_v}) indicates that the current is driven by the entropy gradient and the coefficient is given by 
$D_\alpha/k_{\rm B}$ except the time-dependence.
According to Eq. (\ref{eq:neq2_1}), the entropy gradient increases linearly, then saturates with increasing external field strength, $F$. 
$D_\alpha/k_{\rm B}$ is proportional to $D_\alpha$ given by Eq. (\ref{eq:Dalpha}), and increases with increasing external field strength, while the entropy gradient saturates 
by increasing $F$. 
The entropy gradient is independent of $E_0$ because 
the field dependence is the same between Eq. (\ref{eq:neq2}) for dispersive diffusion and Eq. (\ref{eq:neq1}) for normal diffusion;  
the reason behind could be the pre-averaging with respect to the density of states in the CTRW model. 
Using Eq. (\ref{eq_7}) and $2A/B=2 \tanh[qFb/(2k_{\rm B} T)]/b$, 
we find that the entropy gradient satisfies the fluctuation relation, \cite{Bochkov_77,BOCHKOV_81,Evans_93,Evans_94,Gallavotti_95,Evans_02,Zon_03,Seifert_05,Taniguchi_07,Seifert_12,Chechkin_09,Hurowitz_14,Dieterich_15}
\begin{align}
\frac{G_0(x,t)}{G_0(-x,t)}=\exp \left(\int_0^t dt_1 \frac{d x }{dt_1}\nabla S /k_{\rm B}\right) . 
\label{eq;fl}
\end{align}
By interpreting that $G_0(-x,t)$ is the probability distribution for the time reversed process, 
Eq. (\ref{eq;fl}) indicates that the fluctuation ratio between the time normal and reversed process is governed by the entropy change associated with the fluctuation;  
when the entropy change is positive, the fluctuation of time normal process is more probable than that of time reversed process. 
We point out that the fluctuation relation holds even when 
the entropy gradient given by Eq. (\ref{eq:neq2}) is non-linear function of external field strength ($F$)  
in the CTRW model, where the charge carriers are driven far from equilibrium.  

In Eq. (\ref{eq:neq2}), the ensemble average is taken under an external field ($F$). 
If $ \langle \delta x^2(t) \rangle$ in Eq. (\ref{eq:neq2}) is replaced by the same quantity evaluated in the absence of an external field, 
the upper line in Eq. (\ref{eq:neq2_1}) still holds as pointed out previously. \cite{Barkai_98,Metzler_99,Hou_18}

When the mobility is defined by $\mu(t)=v(t)/(qF)$, using Eq. (\ref{eq:aps_mu3}), 
the mobility thus defined is time dependent, as given by $\mu(t) \propto 1/t^{1-\alpha}$, 
where $\alpha$ is the index of the dispersive diffusion and satisfies $\alpha<1$.  
The mobility coefficient can be defined in a way which is independent of time by $\mu_\alpha=v(t)/[\frac{d}{dt} \int_0^t dt_1 Q(t-t_1)qF ]$. 
Using Eq. (\ref{eq:Dalpha}) and Eq. (\ref{eq:neq2}), we obtain, 
\begin{align}
\mu_\alpha(F)= \frac{\sin (\pi \alpha)}{\pi \alpha} \frac{\left(\gamma_{\rm r}\cosh\left[ q Fb/(2k_{\rm B} T)\right] \right)^\alpha b \tanh[qFb/(2k_{\rm B} T)]}{qF}, 
\label{eq:gmual}
\end{align}
or the form of the generalized Einstein relation, 
\begin{align}
D_\alpha(F)=\mu_\alpha (F) (q F b/2)\coth[qFb/(2k_{\rm B} T)].
\label{eq:gEinsteing}
\end{align}
In the limit of $F \rightarrow 0$, we find, \cite{Metzler_99}
\begin{align}
D_\alpha=\mu_\alpha k_{\rm B} T.
\label{eq:Einstein_al}
\end{align}
The bias dependence in the fluctuation--dissipation relation is obtained using the CTRW model, where the effect of the external electric field is taken into account in the waiting-time distribution.  

\section{Conclusion}

We studied the transient currents and optical absorption using the CTRW model. 
By introducing Arrhenius-type activation of the hopping transition rate with the reduction of the activation barrier by an external electric field, we obtain the field-dependent mobility coefficient and the field-dependent diffusion coefficient for dispersive transport. 
The relation between these transport coefficients reduces to the conventional fluctuation--dissipation relation when diffusion is normal. 
The field dependence of the transport coefficients influences the transit time in the photocurrent kinetics dividing two power-law decay regimes. 

The obtained transient absorption exhibits simple kinetics by varying $AL/B=L\tanh[qFb/(2k_{\rm B} T)]/b\approx qFL/(2k_{\rm B} T)$, whereas the transient currents exhibit rich kinetics.  
Although a transition from $1/t^{1-\alpha}$ to $1/t^{1+\alpha}$ is obtained irrespective of the field strength, when the initial charge carriers are generated only at $x=0$, the kinetics of transient currents depend on the value of $AL/B$ for in-depth carrier injection. 
When $AL/B=1$, the transition from $1/t^{1-\alpha/2}$ to $1/t^{1+\alpha}$ is obtained, whereas the transition from $1/t^{1-\alpha}$ to $1/t^{1+\alpha}$ is obtained for $AL/B>1$. 
The $1/t^{1-\alpha/2}$ time dependence can be interpreted as the spreading of the carrier displacements by dispersive diffusion [Eq. (\ref{eq:Idiff})].
We show that the initial  $1/t^{1-\alpha/2}$ time dependence is obtained for a uniform initial charge distribution and for $\beta$ in the initial distribution less than or equal to $2$ nm$^{-1}$. 
Irrespective of the initial distribution, the long-time decay always follows a $1/t^{1+\alpha}$ time dependence. 
The classical Scher--Montroll theory predicts $a_1+a_2=2$ when the initial photocurrent decay is given by $1/t^{a_1}$ and the asymptotic photocurrent decay is given by $1/t^{a_2}$. 
The results shed light on the interpretation of the power-law exponent in the initial $1/t^{a_1}$ time dependence when $a_1+a_2\neq 2$. 

\newpage
\acknowledgments
This work was supported by JSPS Kakenhi 22K05048. 
T. I. appreciates research grants from Uchida Energy Science Promotion Foundation and Union Tool Scholarship Society.

\section*{AUTHOR DECLARATIONS}
\subsection*{Conflicts of interest}
The authors declare no conflicts of interest.

\subsection*{Data availability}
The data that support the findings of this study are available within the article.

\subsection*{Author contributions}
\noindent
Kazuhiko Seki: Conceptualization (equal); Methodology (supporting); Investigation (lead); Visualization (lead); Writing - original draft (lead). 
Naoya Muramatsu: Methodology (supporting); Resources (supporting); Writing -review \& editing (supporting). 
Tomoaki Miura: Conceptualization (equal); Methodology (equal);  Resources (equal); Writing - review \& editing (supporting). 
Tadaaki Ikoma: Conceptualization (equal); Methodology (equal); Resources (equal);  Writing -review \& editing (supporting).

\section*{Appendix A. Physical interpretation of $\hat{G}_0 (k,s)$ given by Eq. (\ref{eq:invg0s})}
$\hat{G}_0 (k,s)$ in Eq. (\ref{eq:invg0s}) can be expressed as 
\begin{align}
\hat{G}_0 (k,s)=\frac{1-\hat{\psi}_{\rm t} (s) }{s}\sum_{j=0}^\infty \left[\hat{\psi}_{\rm t} (s) \lambda(k) \right]^j.  
\label{eq:seriesp0}
\end{align}
Equation (\ref{eq:seriesp0}) can be understood as follows. 
The factor $\hat{\varphi}_{\rm t} (s)\hat{\psi}_{\rm t} (s)^j$ indicates the Laplace transform of the probability that the carrier executes $j$ hopping transitions at time $t$, where $\hat{\varphi}_{\rm t} (s)=(1-\hat{\psi}_{\rm t} (s))/s$ is the Laplace transform of the remaining probability of the carrier at the trap site without executing further hopping. 
The normalization can be confirmed by 
$\hat{\varphi}_{\rm t} (s)\sum_{j=0}^\infty \hat{\psi}_{\rm t} (s)^j=\hat{\varphi}_{\rm t} (s)/[1-\hat{\psi}_{\rm t} (s)]=1/s$; 
$1/s$ is the Laplace transform of $1$.
$\lambda(k)^j$ indicates the Fourier transform of the displacements caused by $j$ hopping transitions. 
When $j=2$, $\lambda(k)^2$ is the sum of four exponential terms that represent the displacements associated with forward--forward, forward--backward, backward--forward, and backward--backward transitions.

The structure factor $\lambda (k)$ under the bias can be obtained as follows. 
If we denote $\eta(x,j)$ as the probability of just arriving at $x$ after $j$ jumps, we have 
\begin{align}
\eta(x,j)=\Gamma_{\rm p} (F)  \eta(x-b,j-1)+\Gamma_{\rm m} (F) \eta(x+b,j-1),
\label{eq:eta}
\end{align}
which indicates that $\eta(x,j)$ is gained by jumping from $x-b$ with a forward jump probability denoted by $\Gamma_{\rm p}(F)$ and from $x+b$ with a backward jump probability denoted by $\Gamma_{\rm m} (F)$. 
Because jumping should occur with a probability of $1$, we have $\Gamma_{\rm p}+\Gamma_{\rm m}=1$. 

By introducing the Taylor expansion
\begin{align}
\eta(x\pm b,j-1)=\exp\left[\pm b\frac{\partial}{\partial x}
\right]\eta(x,j-1), 
\label{eq:Tylor}
\end{align}
Eq. (\ref{eq:eta}) can be approximated as 
\begin{align}
\eta(x,j)\approx 
\left[1-(\Gamma_p -\Gamma_m)b\frac{\partial}{\partial x} +\frac{b^2}{2} \frac{\partial^2}{\partial x^2} \right]\eta(x,j-1) .
\label{eq:eta1}
\end{align}
Applying the Fourier transformation to both sides of Eq. (\ref{eq:eta1}), we have 
\begin{align}
\eta(k,j)\approx 
\lambda(k)\eta(k,j-1)=\lambda(k)^j ,
\label{eq:eta1F}
\end{align}
where $\lambda(k)$ is given by
\begin{align}
\lambda(k) \approx 1 +iA k-\frac{B}{2} k^2,  
\label{eq:C_app}
\end{align}  
and 
\begin{align}
A&=b\left[\Gamma_p (F)-\Gamma_m(F)\right]=b\tanh[qFb/(2k_{\rm B} T)]\approx qFb^2/(2k_{\rm B} T),
\label{eq:Aapprox_app}\\
B&=b^2
\label{eq:Bapprox_app}
\end{align}
are obtained from Eqs. (\ref{eq:psip})--(\ref{eq:psim}).  
We used the fact that the Fourier transform of the initial distribution $p(x,0)=\delta(x,0)$ is given by $\eta(k,0)=\int_{-\infty}^\infty dx\, \exp(ikx) \delta(x,0)=1$. 
Because each jump is associated with the waiting-time distribution $\hat{\psi}_{\rm t}(s)$ in the Laplace domain, and 
because carriers at $x$ should remain without executing further hopping occurring, we find from Eq. (\ref{eq:eta1F}) that
\begin{align}
\hat{G}_0 (k,s)=\frac{1-\hat{\psi}_{\rm t} (s) }{s}\sum_{j=0}^\infty \left[\hat{\psi}_{\rm t} (s) \lambda(k) \right]^j .  
\label{eq:seriesp0_1}
\end{align}
Equation (\ref{eq:seriesp0_1}) can be rewritten as Eq. (\ref{eq:invg0s}), {\it i.e.},
\begin{align}
\hat{G}_0 (k,s)
&=\frac{1-\hat{\psi }_{\rm t} (s) }{s}\frac{1}{1-\hat{\psi}_{\rm t} (s) \lambda(k)}.   
\label{eq:invg0s_2}
\end{align}

\section*{Appendix B. Derivation of Eq. (\ref{eq:G0_1}) with Eq. (\ref{eq:G0_2}) and Eq. (\ref{eq:f_1})}
Using $1/z=\int_0^\infty du\, \exp(-u z)$, we can express Eq. (\ref{eq:G0}) as \cite{Weiss_94}
\begin{align}
\hat{G}_0 (x,x_{\rm i},s)
&=\frac{1-\hat{\psi }_{\rm t}(s) }{2\pi s}\int_{-\infty}^\infty dk\, \frac{\exp[-ik(x-x_{\rm i})]}
{1-\hat{\psi }_{\rm t}(s) \lambda(k)}, 
\nonumber \\
&=\frac{1-\hat{\psi }_{\rm t}(s) }{2\pi s}
\int_0^\infty du\,
\int_{-\infty}^\infty dk \exp
\left[
-ik(x-x_{\rm i})-\left(1-\hat{\psi}_{\rm t}\lambda\right)u
\right]
\label{eq:g0_1}\\
&=\frac{1-\hat{\psi }_{\rm t}(s) }{s\sqrt{2\pi B\hat{\psi}_{\rm t}}}
\int_0^\infty \frac{du}{\sqrt{u}}\,
\exp
\left[-\left(1-\hat{\psi}_{\rm t}\right)u-
\frac{\left(x-x_{\rm i}-A\hat{\psi}_{\rm t}u\right)^2}{2 B \hat{\psi}_{\rm t} u}
\right]
\label{eq:g0_12}\\
&=\frac{1-\hat{\psi }_{\rm t} }{s\sqrt{\hat{\psi}_{\rm t}\left[2 B\left(1- \hat{\psi}_{\rm t}\right)+A^2 \hat{\psi}_{\rm t}
\right]}}
\exp
\left[\frac{A}{B}
\left(x-x_{\rm i}-|x-x_{\rm i}|
\sqrt{1+
\frac{2B\left(1- \hat{\psi}_{\rm t}\right)}{A^2\hat{\psi}_{\rm t}
}
}
\right) 
\right] . 
\label{eq:g0_2}
\end{align}
From Eq. (\ref{eq:f}), we obtain
\begin{align}
\hat{f}(L, x_{\rm i}, s)=\exp
\left[\frac{A}{B}
\left(L-x_{\rm i}-|L-x_{\rm i}|
\sqrt{1+
\frac{2B\left(1- \hat{\psi}_{\rm t}\right)}{A^2\hat{\psi}_{\rm t}
}
}
\right) 
\right] . 
\label{eq:f1}
\end{align}

\section*{Appendix C. Derivation of Eq. (\ref{eq:constePot})}
The constant voltage difference can be expressed using the surface charge on the metal electrode at $x=0$ (back contact) denoted by $q_0(t)$, and the surface charge on the metal electrode at $x=L$ (front contact) denoted by $q_L(t)$ [\onlinecite{Hirao_95}]. 
$q_0(t)$ is supplied from the electrode and $q_L(t)$ is supplied from the electrode as well as from the sample by electron transfer when the reflecting boundary condition is imposed at $x=0$.  
Using the Gauss theorem, we can express the electric field as \cite{Hirao_95,Nishizawa_06}
\begin{align}
\epsilon \epsilon_0 \frac{d}{dx} E(x,t)=\rho(x,t)+q_0(t) \delta (x-\epsilon_\delta) + q_L(t) \delta(x-L+\epsilon_\delta), \mbox{ for } 0\leq x \leq L
\label{eq:GaussHirano}
\end{align}
when the sample is placed between the metal electrodes ($E=0$ for $x\leq 0$ and $x\geq L$)
with the relative permittivity ($\epsilon$), where $\epsilon_0$ indicates the permittivity of vacuum. 
$\epsilon_\delta$ is the smallest length scale; later, we take the limit of $\epsilon_\delta \rightarrow 0$. 
We obtain
\begin{align}
\epsilon \epsilon_0 E(x,t)=\int_0^x dx_1 \rho(x_1,t)+q_0 (t)+ \int_0^x dx_1 q_L(t) \delta(x_1-L+\epsilon_\delta) .
\label{eq:GaussHiranoint}
\end{align}
By further integrating with respect to $x$, we can obtain the potential difference by
\begin{align}
\Delta V= \int_0^L E(x) dx &= \frac{1}{\epsilon \epsilon_0} \left[\int_0^L dx (L-x) \rho(x,t) +q_0(t) L +\int_0^L dx (L-x) q_L(t) \delta(x-L+\epsilon_\delta)\right],
\label{eq:pot}\\
&= \frac{1}{\epsilon \epsilon_0} \left[\int_0^L dx (L-x) \rho(x,t) +q_0(t) L \right],
\label{eq:pot1}
\end{align}
where we have used
\begin{align}
\int_0^L dx \int_0^x dx_1 \rho(x_1,t) =\int_0^L dx (L-x) \rho(x,t) 
 \label{eq:potc1}
\end{align}
and the limit of $\epsilon_\delta \rightarrow 0$. 
Under the constant potential difference, we finally obtain \cite{Nishizawa_06}
\begin{align}
J(t)=\frac{d}{dt} q_0(t)
&=-\frac{1}{L}\frac{d}{dt}  \int_0^L dx(L-x) \rho(x,t) .
\label{eq:dQ_21}
\end{align}
By introducing $\rho(x,t)=q\int_0^L d x_{\rm i}\, G (x,x_{\rm i},t)p_{\rm i} (x_{\rm i})$, where $G (x,x_{\rm i},t)$ is given by Eq. (\ref{eq:SM}), we obtain Eq. (\ref{eq:constePot}). 

\section*{Appendix D. The Green function satisfying the reflecting boundary condition at $x=0$}
In the absence of a boundary at $x=0$ and $x=L$, the Laplace transform of $G_0(x,x_{\rm i},t)$ (Eq. (\ref{eq:G0}) is given by 
\begin{align}
\hat{G}_0(x,x_{\rm i},s)
&= \frac{1-\hat{\psi}_{\rm t}}{s}\frac{1}{2\pi} \int_{-\infty}^\infty dk\, \frac{\exp[-ik(x-x_{\rm i})]}{1-\hat{\psi}_{\rm t} \left[1+iAk-(B/2)k^2 \right]},
\label{eq:aps_1_2}
\end{align}
where $A$ and $B$ indicate the expansion coefficient given by $\lambda(k)\approx 1 + iAk-Bk^2/2$ and $x_{\rm i}$ is the initial position.

By introducing $k_1=k-iA/B$, we find 
\begin{align}
-iAk+(B/2)k^2 =A^2/(2B)+(B/2) k_1^2, 
\label{eq:aps_5}
\end{align}
and Eq. (\ref{eq:aps_1_2}) can be rewritten as
\begin{align}
\hat{G}_0(x,x_{\rm i},s)&= \frac{1-\hat{\psi}_{\rm t}}{s}\frac{1}{2\pi} \int_{-\infty}^\infty dk_1\, \frac{\exp[-i(k_1+iA/B)(x-x_{\rm i})]}{1-\hat{\psi}_{\rm t} \left[1+A^2/(2B)+(B/2) k_1^2 \right]} .
\label{eq:aps_6_2}
\end{align}
Therefore, we obtain
\begin{align}
G_0(x,x_{\rm i},t)=\exp\left[(A/B)(x-x_0)\right]g_0(|x-x_{\rm i}|,t),  
\label{eq_7}
\end{align}
where $g_0(|x-x_{\rm i}|,t)$ is obtained by the inverse Laplace transform of $\hat{g}_0(x-x_{\rm i},s)$;  
$\hat{g}_0(x-x_{\rm i},s)$  is given by
\begin{align}
\hat{g}_0(x-x_{\rm i},s)&= \frac{1-\hat{\psi}_{\rm t}}{s}\frac{1}{2\pi} \int_{-\infty}^\infty dk\, \frac{\exp[-ik(x-x_{\rm i})]}{1-\hat{\psi}_{\rm t} \left[1+A^2/(2B)+(B/2) k^2 \right]} ,
\label{eq:aps_8}
\end{align}
where $\hat{g}_0(x-x_{\rm i},s)$ is even function of $x-x_{\rm i}$.

Now, we consider the reflecting boundary condition at $x=0$. 
The solution can be expressed as 
\begin{align}
\hat{G}_{\rm r}(x,x_i,s)=\exp\left(\frac{A}{B} (x-x_i)\right) \hat{g}_{\rm r}(x,x_i,s) . 
\label{eq:refl0}
\end{align}
The reflecting boundary condition can be expressed as 
\begin{align}
\left. AG_{\rm r}(0,x_{\rm i},t) -\frac{B}{2} \frac{\partial}{\partial x}G_{\rm r}(x,x_{\rm i},t) \right|_{x=0}=0
\label{eq:aps_10_1_0}
\end{align}
and 
\begin{align}
\left. A g_{\rm r}(0,x_{\rm i},t) -B \frac{\partial}{\partial x}g_{\rm r}(x,x_{\rm i},t) \right|_{x=0}=0 .
\label{eq:aps_10_1}
\end{align}
We express $g_{\rm r}(x,x_{\rm i},t)$ by superposition of $g_0(x-x_{\rm i},t)$ as  
\begin{align}
g_{\rm r}(x,x_{\rm i},t)=g_0(x-x_{\rm i},t)+g_0(x+x_{\rm i},t)+\int_0^\infty d\xi g_0(x+x_{\rm i}+\xi,t)F(\xi),
\label{eq:aps_11}
\end{align}
where $F(\xi)$ are determined from the boundary condition. 
Notably,
\begin{align}
\left. \frac{d}{dx}g_0(x\pm x_{\rm i},t)\right|_{x=0}&=\frac{1-\hat{\psi}_{\rm t}}{s}\frac{1}{2\pi} \int_{-\infty}^\infty dk\, \frac{(-ik)\exp(\mp ik x_{\rm i})}{1-\hat{\psi}_{\rm t} \left[1+A^2/(2B)+(B/2) k^2 \right]} 
\label{eq:12_1}\\
&=\frac{1-\hat{\psi}_{\rm t}}{s}\frac{1}{2\pi} \int_{-\infty}^\infty dk\, \frac{(\mp ik)\exp(- ik x_{\rm i})}{1-\hat{\psi}_{\rm t} \left[1+A^2/(2B)+(B/2) k^2 \right]} ,
\label{eq:12_2}
\end{align}
which leads to 
\begin{align}
\left. \frac{d}{dx}g_0(x+ x_{\rm i},t)\right|_{x=0}&=-\left. \frac{d}{dx}g_0(x- x_{\rm i},t)\right|_{x=0} .
\label{eq:13}
\end{align}
Therefore, if Eq. (\ref{eq:aps_11}) is substituted into Eq. (\ref{eq:aps_10_1}), the first derivatives of the first two terms in Eq. (\ref{eq:aps_11}) cancel each other. 
Using partial integration, we evaluate the rest of the first derivatives as  
\begin{align}
\frac{d}{dx} \int_0^\infty &d\xi g_0(x+x_{\rm i}+\xi,t)F(\xi)=-\int_0^\infty d\xi \frac{d}{d \xi}g_0(x+x_{\rm i}+\xi,t)F(\xi)
\label{eq:aps_14_1}\\
&=-g_0(x+x_{\rm i},t)F(0)-\int_0^\infty d\xi g_0(x+x_{\rm i}+\xi,t)\frac{d}{d \xi} F(\xi). 
\label{eq:aps_14_2}
\end{align}
By substituting Eq. (\ref{eq:aps_14_2}) into Eq. (\ref{eq:aps_10_1}), we obtain 
\begin{align}
\frac{d}{d \xi} F(\xi)=-\frac{A}{B} F(\xi)
\label{eq:15_1}
\end{align}
with $F(0)=-2A/B$. 
The solution is $F(\xi)=-(2A/B) \exp\left[-(A/B) \xi \right]$.
Therefore, we obtain
\begin{align}
g_{\rm r}(x,x_{\rm i},t)=g_0(x-x_{\rm i},t)+g_0(x+x_{\rm i},t)-\frac{2A}{B} \int_0^\infty d\xi g_0(x+x_{\rm i}+\xi,t)\exp\left(-\frac{A}{B} \xi \right).
\label{eq:aps_16}
\end{align}
The last integral can be evaluated using Eqs. (\ref{eq:g0_2}) and (\ref{eq_7}) as 
\begin{multline}
\frac{A}{B}\int_0^\infty d\xi \hat{g}_0(x+x_{\rm i}+\xi,s)\exp\left(-\frac{A}{B} \xi \right)
=
\\
\frac{\left[1-\hat{\psi }_{\rm t}\right]\exp\left[-(A/B)(x+x_{\rm i})\right] }{s\sqrt{\hat{\psi}_{\rm t}\left[2 B\left(1- \hat{\psi}_{\rm t}\right)+A^2 \hat{\psi}_{\rm t}
\right]}}
\left[1+\sqrt{1+
\frac{2B\left(1- \hat{\psi}_{\rm t}\right)}{A^2\hat{\psi}_{\rm t}
}}
\right]^{-1} . 
\label{eq:aps_17}
\end{multline}

When the reflecting boundary condition is imposed at $x=0$, Eqs. (\ref{eq:SM}) and (\ref{eq:constePot}), which take into account the charge extraction at $x=L$, still hold. 
In Eqs. (\ref{eq:SMs}) and (\ref{eq:f}), $\hat{G}_0 (x,x_{\rm i},s)$ should be replaced with $\hat{G}_{\rm r}(x,x_{\rm i},s)$; $\hat{G}_{\rm r}(x,x_{\rm i},s)$ is given by Eq. (\ref{eq:refl0}) obtained under the reflecting boundary condition at the back contact ($x=0$) under an external field, \cite{TYUTNEV_15} where $\hat{g}_{\rm r}(x,x_i,s)$ is given by Eq. (\ref{eq:aps_16}).
In this manner, the reflecting boundary condition at $x=0$ can be set in addition to the boundary condition at $x=L$. 
However, the photocurrent obtained without imposing the boundary condition even at  $x=L$ turned out to be sufficient to study the power-law kinetics; for brevity, we do not impose the additional boundary condition at $x=0$.

\section*{Appendix E. Derivation of Eq. (\ref{eq:Idiff_1})}
Using the Laplace transform of Eq. (\ref{eq:constePot}), which defines the current density, we obtain
\begin{align}
\hat{J}(s)&=-\frac{qs}{L^2} \int_0^L dx \int_0^L dx_{\rm i} (L-x)\hat{G} (x,x_{\rm i},s) , 
\label{eq:constePot_s_1}\\
&\approx-\frac{qs}{L^2} \int_0^L dx \int_0^x dx_{\rm i} (L-x)\hat{G} (x,x_{\rm i},s) , 
\label{eq:constePot_s_2}\\
&\approx\frac{qB\left(1-\hat{\psi}_{\rm t}\right)}{2 A^2 + 4B\left(1-\hat{\psi}_{\rm t}\right)-2A \sqrt{A^2+2B\left(1-\hat{\psi}_{\rm t}\right)}}  ,
\label{eq:constePot_s_3}
\end{align}
where we assumed that the sample thickness is greater than the hopping distance ($L>b$) and that $qFb/(k_{\rm B} T)<1$ to derive the last line. 
Because we are interested in the kinetics at early times, we consider the case $A^2/(2B)<1-\hat{\psi}_{\rm t}(s)$.
In this case, $\hat{J}(s)$ can be further simplified as 
\begin{align}
\hat{J}(s)&\approx\frac{1}{4-2A \sqrt{2/\left[B\left(1-\hat{\psi}_{\rm t}\right)\right]}} 
\label{eq:constePot_s_4}\\
&\approx\frac{1}{4} \left(1+\frac{A }{2}\sqrt{\frac{2}{B\left(1-\hat{\psi}_{\rm t}\right)}}\right)
\label{eq:constePot_s_5}\\
&\approx\frac{A }{8}\sqrt{\frac{2}{B\left(1-\hat{\psi}_{\rm t}\right)}} .
\label{eq:constePot_s_6}
\end{align}
The inverse Laplace transform of Eq. (\ref{eq:constePot_s_6}) using Eq. (\ref{eq:psist1}) yields Eq. (\ref{eq:Idiff_1}). 

%


\end{document}